\documentclass[preprint,showpacs,eqsecnum,aps]{revtex4}
\usepackage{graphicx}
\usepackage{epsfig}
\begin{document}

\title{A solvable model which has X(5) as a limiting symmetry and removes some inherent drawbacks} 

\author{A. A. Raduta$^{a),b)}$, A. C. Gheorghe$^{(b)}$, P. Buganu$^{b)}$ and Amand Faessler$^{c)}$}

\address{$^{b)}$Department of Theoretical Physics and Mathematics,Bucharest University, POBox MG11, Romania}

\address{$^{a)}$Institute of Physics and Nuclear Engineering, Bucharest, POBox MG6, Romania}

\address{$^{c)}$Institut fuer Theoretische Physik, der Universitaet Tuebingen,
auf der Morgenstelle 14, Germany}

\begin{abstract}
Solvable Hamiltonians for the $\beta$ and $\gamma$ intrinsic shape coordinates are proposed.  The eigenfunctions of the $\gamma$  Hamiltonian are  spheroidal periodic functions,  while  the Hamiltonian for the $\beta$  degree of freedom involves the Davidson's potential and admits  eigenfunctions which can be expressed in terms of the generalized Legendre polynomials. The proposed model goes to $X(5)$ in the limit of $|\gamma|$-small. Some drawbacks of the X(5) model, as are the eigenfunction periodicity and the $\gamma$ Hamiltonian hermiticity, are absent in the present approach.  Results of numerical applications to $^{150}$Nd, $^{154}$Gd and $^{192}$Os are in good agreement to the experimental data.  Comparison with $X(5)$ calculations suggests that the present approach provides a quantitative better description of the data. This is especially true for the excitation energies in the gamma band.     
\end{abstract}

\pacs{: 21.10.Re, 23.20.Lv, 21.60. Ev}
\maketitle

\renewcommand{\theequation}{1.\arabic{equation}}
\setcounter{equation}{0}
\section{Introduction}
\label{sec:level1}

Since the liquid drop model was developed \cite{Bohr}, the
quadrupole shape coordinates  were widely used  by both
phenomenological and microscopic formalisms to describe the basic properties of
nuclear systems. Based on these coordinates, one defines quadrupole
boson operators in terms of which model Hamiltonians and transition operators
are defined. Since the original spherical harmonic liquid drop model was able
to describe only a small amount of data for spherical nuclei, several
improvements have been added. Thus, the Bohr-Mottelson model was generalized by
Faessler and Greiner \cite{GrFa}
in order to describe the small oscillations around a deformed shape which
results in obtaining a flexible model, called vibration rotation model,
suitable for the description of deformed nuclei. Later on \cite{Gneus} this picture was
extended by including anharmonicities as low order invariant  polynomials in the
quadrupole coordinates. With a suitable choice of the parameters involved in the model
Hamiltonian the equipotential
energy surface may exhibit several types of minima \cite{Hess} like spherical,
deformed prolate, deformed oblate, deformed triaxial, etc.
 To each equilibrium shape, specific properties for excitation energies and electromagnetic transition
 probabilities show up. Due to this reason,
one customarily says that static values of intrinsic coordinates determine a phase for the
nuclear system. A weak point of the boson description with a complex anharmonic Hamiltonian consists of the large number of the structure parameters which are
to be fitted.  A much smaller number of parameters is used by the
coherent state model (CSM) \cite{Rad1} which uses a restricted collective space generated through angular momentum projection by three deformed orthogonal functions of coherent type. The model is able to describe in a realistic fashion transitional and well deformed nuclei of various shapes including states of high and very high angular momentum. Various extensions to include other degrees of freedom like isospin
\cite{Rad2}, single particle \cite{Rad3} or octupole \cite{Rad4} degrees of freedom have been formulated \cite{Rad5}.
  
It has been noticed that a given nuclear phase may be
associated to a certain symmetry. Hence, its properties may be described with
the help of the irreducible representation of the respective symmetry group.
Thus, the gamma unstable nuclei can be described by the $O(6)$ symmetry
\cite{Jean}, the symmetric rotor by the $SU(3)$ symmetry and the spherical vibrator by the
$U(5)$ symmetry.
The gamma triaxial nuclei are characterized  by the invariance symmetry group $D2$ 
\cite{Filip} of the rigid triaxial rotor Hamiltonian.
Thus, even in the 50's, the   symmetry properties have been greatly appreciated.
 However, a big push forward was brought by the interacting boson
 approximation
(IBA) \cite{Iache,Iache1}, which succeeded to describe the basic properties of a large number of
nuclei in terms of the symmetries associated to a system of quadrupole (d) and
monopole (s) bosons  which  generate a $U(6)$ algebra. The three
limiting symmetries $U(5)$, $O(6)$, $SU(3)$, are dynamic symmetries
for $U(6)$. Moreover, for each of these symmetries a specific group reduction
chain  provides the quantum numbers characterizing the states, which are suitable
  for a certain region of nuclei. Besides  the virtue of unifying the group
  theoretical descriptions of nuclei exhibiting different symmetries,
  the procedure defines very simple reference pictures for the limiting cases. For nuclei
 lying close to the region characterized by a certain symmetry,
 the perturbative corrections are to be included.

In Ref. \cite{Cast}, a new classification scheme was provided,  all
nuclei being distributed within a symmetry triangle. The vertices of this triangle symbolize the
$U(5)$ (vibrator), $O(6)$ (gamma soft) and $SU(3)$ (symmetric rotor), while the legs
of the triangle denote the transitional region. Properties of nuclei lying far
from vertices are difficult to be explained since the
states have some characteristics of one vertex while some others are easy to be
described by using the adjacent symmetry. The transition from one phase to another reaches a critical point depending on the specific parametrization as well as on the transition type 
\cite{Row1,Row2,Row3}. In Ref. \cite{Gino,Diep}, it has been proved
that on the $U(5)-O(6)$ transition leg there exists a critical point
for a second order phase transition while the
$U(5)-SU(3)$ leg has a first order phase transition. 

Recently, Iachello \cite{Iache2,Iache9} pointed out that these critical points
correspond to
distinct symmetries, namely $E(5)$ and $X(5)$, respectively. For the critical value
of an
ordering parameter, energies are given by the zeros of a Bessel function of half integer and irrational indices, respectively.
The description of low lying states in terms of Bessel functions was used first by Jean and Willet \cite{Jean}, but the interesting feature saying that this is a critical picture in a phase transition and defines a new symmetry, was indeed advanced first in Ref.\cite{Iache2}. 

Representatives for
the two symmetries have been experimentally identified. To give an example, the relevant data for $^{134}$Ba \cite{Zam}
and $^{152}$Sm \cite{Zam1}
suggest that they are close to the $E(5)$ and $X(5)$ symmetries, respectively.
Another candidate for E(5) symmetry, proposed by Zamfir {\it et al.} \cite{Zam2}
is $^{102}$Pd. Using a simple IBA Hamiltonian, in Ref.\cite{Da-li}, the low lying
spectrum of
$^{108}$Pd is realistically described. Comparing the E(5) predictions with the
experimental data concerning energy ratios in the ground band and the normalized
E2 transition probabilities for the states $4^+$ and $0^+_2$,
one concludes that this nucleus is a good E(5) candidate.
However, in order to decide which  Pd isotope is closer to an E(5) behavior, further
investigations are necessary.
A systematic search for $E(5)$ behavior in nuclei has been reported in Ref.\cite{Clar}.

Short after the pioneering papers concerning critical point  symmetries
appeared, some other attempts have
been performed, using other potentials like Coulomb, Kratzer \cite{Fort} and
Davidson potentials \cite{Bona}. These potentials yield also Schr\"{o}dinger
solvable equations and the corresponding results may be interpreted in terms 
of symmetry groups.

In Ref.\cite{Rad05} we advanced the hypothesis that the critical point in a phase transition is state dependent. We tested this
with a hybrid model for $^{134}$Ba and $^{104}$Ru.

The departure from the gamma unstable picture has been treated by several authors whose contributions are reviewed by Fortunato in Ref.\cite{Fortu2}. The difficulty in treating the gamma degree of freedom consists in the fact that this is coupled to the rotational degrees of freedom. A full solution for the Bohr-Mottelson Hamiltonian including an explicit treatment of gamma deformation can be found in Refs.\cite{Ghe78,Rad78}. Therein, we treated separately also the gamma unstable and the rotor Hamiltonian. A more complete study of the rotor Hamiltonian and the distinct phases associated to a tilted moving rotor is given in Ref. \cite{Rad98}.
Distinct solutions, expressed in laboratory frame shape coordinates, have been reported in Refs.
\cite{Moshy,Willi,Row}. The gamma dependent part of the wave function has been found as a solution of a specific differential equation in Ref.\cite{Bes}.  

Finding  the $\gamma$ depending part of the wave function becomes even more complicated when we add to the liquid drop Hamiltonian a potential depending on $\beta$ and $\gamma$ at a time.
 To simplify the starting problem related to the inclusion of $\gamma$ one uses  model potentials which are sums of a $\beta$ term, $V(\beta)$, and a factorized $\beta-\gamma$ term $U(\gamma)/\beta^2$. In this way the nice feature for the beta variable to be decoupled from the remaining four variables, specific to the harmonic liquid drop, is preserved. In the next step, the potential in gamma is expanded  around $\gamma=0$ or $\gamma=\frac{\pi}{6}$.
In the first case if only the singular term is retained one obtains the infinite square well model described by Bessel functions in $\gamma$. If the $\gamma^2$ term is added to this term, the Laguerre functions are the eigenstates of the approximated $\gamma$ depending Hamiltonian, which results in defining the so called X(5) approach. 

Note that any approximation applied to the $\gamma$-Hamiltonian modifies automatically the differential equation for $\beta$. Indeed, the centrifugal term $\tau(\tau+3)/\beta^2$ disappears but another one is expected to come from the $\beta-\gamma$ coupling after some approximations are performed.

The drawback of these approximations consist in that the resulting functions are not periodic, as the starting Hamiltonian is. Moreover, they are orthonormalized on unbound intervals although the underlying equation was derived under the condition of $|\gamma|$ small. Moreover, the scalar product of the resulting functions is not defined with the integration measure $|\sin3\gamma| d\gamma$ as  happens in the liquid drop model. Under these circumstances it happens that the approximated Hamiltonian in 
$\gamma$ looses its hermiticity.

In a previous short publication \cite{Rad07} we proposed a scheme where the $\gamma$ Hamiltonian is solvable Hamiltonian. Moreover, its eigenstates are the spheroidal functions, which are periodic. Also one proves that the X(5) model is recovered in the limit of $\gamma$-small. Here we give details about the calculations and describe some new numerical applications. Moreover, we complete our formalism by considering a Schr\"{o}dinger equation in the beta coordinate, involving a potential of the Davidson's type.

By the numerical applications  we want to see whether curing the mentioned drawbacks for gamma  wave functions and Hamiltonian, would bring substantial quantitative corrections to energies and E2 transitions given by the $X(5)$ formalism. 
Even if the found corrections are not dramatic the  virtue of the proposed formalism to extend the $X(5)$ description to a theory which is consistent with the symmetries of the starting Hamiltonian as well with some basic principles of Quantum Mechanics concerning  the hermiticity property of the model Hamiltonian remains an important achievement of the present paper.

The above objectives are reached according to the following plan. The starting Hamiltonian is presented in Section II, where the separability conditions are discussed. Several methods for treating $\beta$ are described in Subsection A while 
the Hamiltonian in  $\gamma$ is studied in Subsection B.
The specific procedure of the present paper is presented in Section III. Numerical applications are given in Section IV.
The final conclusions are summarized in Section V.

\renewcommand{\theequation}{2.\arabic{equation}}
\setcounter{equation}{0}
\section{The starting Hamiltonian}
\label{sec:level2}

Written in the intrinsic frame of
reference, the original Bohr-Mottelson Hamiltonian has the expression:
\begin{equation}
H=-\frac{\hbar^2}{2B}\left[\frac{1}{\beta^4}\frac{\partial }{\partial \beta}
\beta^4 \frac{\partial }{\partial \beta}+\frac{1}{\beta^2\sin {3\gamma} }
\frac{\partial}{\partial \gamma}\sin{3\gamma}\frac{\partial}{\partial \gamma}
-\frac{1}{4\beta^2}\sum_{k=1,2,3}\frac{Q_k^2}{\sin^2(\gamma-\frac{2}{3}\pi k)}
\right]+V(\beta,\gamma),
\label{Has}
\end{equation}
where the dynamic deformation variables are denoted by $\beta$ and $\gamma$
while the intrinsic angular momentum components by $Q_k$, with $k=1,2,3$.
Within the liquid drop model the potential energy depends quadratically on $\beta$. Here we assume that the potential energy depends on both deformation variables, beta and gamma.
Without exception the solvable models proposed for a simultaneous description of $\beta$ and $\gamma$ variables adopt the variable separation methods. Two situations are to be distinguished:

a) If the potential energy term is depending  on deformation variables in a separable manner:
\begin{equation}
V(\beta,\gamma)=V(\beta)+U(\gamma),
\end{equation}
and some additional assumptions are adopted, the eigenvalue equation associated to $H$ (\ref{Has}) can be separated in 
two parts, one
equation describing the beta variable and the other one the gamma
deformation and the Euler angles
$\Omega=(\theta_1,\theta_2,\theta_3)$. Indeed, in this case the
separation of variables, achieved by various models, is based on two approximations \cite{Caprio,Bona3}: i) restriction to small values of $\gamma$, i. e. $|\gamma|\ll 1$; 
ii) replacing the factor $1/\beta^2$ by $1/\langle \beta^2\rangle$ in the terms involved in the equation for $\gamma$.
The diagonalization of the Bohr-Mottelson Hamiltonian shows that the first approximation is valid for large $\gamma$ stiffness while the second one for small $\gamma$ stiffness \cite{Bona3}.

b) A complete separation of equations for the two variables, $\beta$ and $\gamma$ is possible if we choose the potential
\begin{equation}
V(\beta,\gamma)=V(\beta)+U(\gamma)/\beta^2.
\end{equation}
In this way the  approximation consisting of replacing $\beta^2$ by $\langle\beta^2\rangle$ is avoided but the restriction to the case  $|\gamma|\ll 1$ is still kept. Thus, the theory involves one parameter which is the $\gamma$ stiffness which affects the excitation energies in both the beta and gamma bands.
 
In what follows the two equations, for $\beta$ and $\gamma$,  will be considered separately.

\subsection{The treatment of the $\beta $  Hamiltonian}

The  solvable models for $\beta$, presented here, have been used by E(5) formalisms, which ignore the potential in $\gamma$. Considering the potential in $\gamma$, of course, the picture for $\beta$ is changed. However, as we shall see later on, it is very easy to derive analytically the energies and wave functions associated to $\beta$ from the corresponding results of the E(5) descriptions. Actually, this is the motivation for reviewing, here, the beta solvable models.

 The equation in $\beta$ is:
\begin{equation}
\left[-\frac{1}{\beta^4}\frac{\partial }{\partial \beta}\beta^4
\frac{\partial }{\partial \beta}+\frac{\Lambda}{\beta^2}+u(\beta)\right ]f(\beta)
=\epsilon f(\beta),
\label{epsf}
\end{equation}
where   $ \Lambda$ is the eigenvalue of the Casimir operator of the $SO(5)$ group.
This is related with the seniority quantum number $\tau$,  by $\Lambda=\tau(\tau+3)$.
The 'reduced' potential $u(\beta)$  and energy $\epsilon$ are defined as:
\begin{equation}
E = \frac{\hbar ^2}{2B}\epsilon,\;\;\;V= \frac{\hbar ^2}{2B}u.
\label{EU}
\end{equation}
where $E$ denotes the eigenvalue of the Hamiltonian $H$ corresponding to the potential $V(\beta)$.
Here we mention the most used potentials for $\beta$:

\vskip0.3cm
\subsubsection{The case of $u(\beta)=\beta^2$ }
\vskip0.3cm

A full description of the eigenstates of the Bohr-Mottelson Hamiltonian
satisfying the symmetry $U(5)\supset SO(5)\supset SO(3)\supset SO(2)$,
may be found in Refs.\cite{Ghe78}. In particular, the solution of the radial equation
(\ref{epsf}) with $u(\beta)=\beta^2$ is easily obtained by bringing first  Eq.(
\ref{epsf}) to the
standard Schr\"{o}dinger form by changing the function f to $\psi$ by:
\begin{equation}
\psi(\beta)=\beta^2 f(\beta).
\label{psi}
\end{equation}
The equation obeyed by the new function $\psi$, is:
\begin{equation}
\frac{d^2{\psi}}{d\beta^2}+
\left[\epsilon-\beta^2-
\frac{(\tau+1)(\tau+2)}{\beta^2}\right]\psi=0.
\label{Sch}
\end{equation}
This equation is analytically solvable. The solution is:
\begin{eqnarray}
\psi_{n\tau}(\beta)&=&\sqrt{\frac{2(n!)}{\Gamma(n+\tau+5/2)}}
L^{\tau+3/2}_n(\beta^2)\beta^{\tau+2}\exp(-\beta^2/2), \\
\epsilon_n&=&2n+\tau+5/2,\;\;\;n=0,\;1,\;2,\;...;\tau=0,1,2,3,...
\label{spectr}
\end{eqnarray}
where $L_n^{\nu}$ denotes the generalized Laguerre polynomials. The number of
polynomial nodes
is denoted by $n$ and is related to the number of the quadrupole bosons ($N$) in
the state, by: $N=2n+\tau$.
Consequently, the initial equation (\ref{epsf}) has the solution
\begin{equation}
f_{n\tau}=\beta^{-2}\psi_{n\tau}\;.
\end{equation}

The spectrum, given by Eq. (\ref{spectr}), may be also obtained  by using the unitary representation of
the $SU(1,1)$ group with
the Bargman index $k=(\tau+5/2)/2.$  Indeed, the standard generator for $SU(1,1)$ are:
\begin{equation}
K_0=\frac{1}{4}H_0,\;\;\;\;K_{\pm}=\frac{1}{4}\left[\frac{(\tau+1)(\tau+2)}{\beta^2}-(\beta\pm\frac{d}{d\beta})^2\right],
\label{su11}
\end{equation}
where
\begin{eqnarray}
H_0&=&-\frac{d^2}{d\beta^2}+\beta^2 +\frac{(\tau+1)(\tau+2)}{\beta^2},
\nonumber\\
H_0\psi_n &=& \epsilon_n\psi_n.
\label{H0}
\end{eqnarray}
$H_0$ obeys the following equations:
\begin{equation}
\left[K_-,K_+\right] = -\frac{1}{2}H_0,\;\;
\left[K_{\pm},H_0\right] = \pm 4K_{\pm}.
\label{comut}
\end{equation}
\vskip0.3cm
\subsubsection{ Davidson's potential}
\vskip0.3cm

Another potential in $\beta$ which yields a solvable model is due to Davidson \cite{Dav}:

\begin{equation}
u(\beta )=\beta ^{2}+\frac{\beta_0^4}{\beta ^2}.
\label{Davi}
\end{equation}
This potential has been used by several authors in different contexts \cite{Elli,Row,Bona}.
The potential has been used by Bonatsos {\it at al} \cite{Bona3}, to describe the dynamic deformation variable $\beta$.
For this potential the above equations (2.6-2.10) hold for $\tau $
replaced \cite{Bona} by 
\begin{equation}
\tau ^{\prime }=-\frac{3}{2}+\left[ \left( \tau +\frac{3}{2}\right)
^{2}+\beta_0^4\right] ^{1/2}.  \label{dav0}
\end{equation}
\vskip0.3cm
In particular, the excitation energies have the expressions:
\begin{equation}
E_{n,\tau}=2n+1+\left[ \left( \tau +\frac{3}{2}\right)
^{2}+\beta_0^4\right] ^{1/2}.
\end{equation}
The factor $\beta_0^4$ is considered to be a free parameter which is to be determined variationally for each angular momentum, as suggested in
Ref. \cite{Bona}:
\begin{equation}
\frac{d^2R^{(g)}_L}{d\beta_0^2}=0,
\end{equation}
where $R$ denotes the ratio of the excitation energy of the ground band state  $L^+$
and the excitation energy of the state $2^+_g$.

\subsubsection{Five dimensional infinite well}

Now, let us turn our attention to the situation considered by Iachello in
Ref.\cite{Iache2}, where
the potential term associated to the spherical to gamma unstable shape transition
 is so flat that it can be mocked up as a infinity square well
\begin{eqnarray}
u(\beta)=\left\{\matrix{0,&\beta\leq \beta_w,\cr
                       \infty, & \beta > \beta_w.}\right .
\label{udeb}
\end{eqnarray}
A more convenient form for the equation in $\beta$, is obtained through the function
transformation:
\begin{equation}
\varphi(\beta)=\beta^{3/2}f(\beta),
\label{fi}
\end{equation}
The equation for $\varphi$ is
\begin{equation}
\frac{d^2\varphi}{d\beta^2}+\frac{1}{\beta}\frac{d\varphi}{d\beta}+
\left[\epsilon-u(\beta)-\frac{(\tau +3/2)^2}{\beta ^2}\right]\varphi=0.
\label{diffi}
\end{equation}
Changing the variable $\beta$ to $z$ by
\begin{equation}
z=k\beta ,\;\;\; k=\sqrt{\epsilon}
\label{zsik}
\end{equation}
and denoting with $\widetilde {\varphi}(z)=\varphi(\beta)$  the function of the new
variable, one arrives at:
\begin{equation}
\frac{d^2\widetilde{\varphi}}{dz^2}+\frac{1}{z}\frac{d\widetilde{\varphi}}{dz}
+\left[1-
\frac{(\tau+3/2)^2}{z^2}\right]\widetilde{\varphi}=0.
\label{fitild}
\end{equation}
This equation is analytically solvable, the solutions being the Bessel
functions of half
integer order, $J_{\tau+3/2}(z)$. Since for $\beta>\beta_w$ the function
$\widetilde{\varphi}$ is equal to zero, the continuity condition requires that
the solution inside the well must vanish for the value of $\beta$ equal to $\beta_w$.
This, in fact, yields a quantized form for the eigenvalue $E$. Indeed,
let $x_{\xi,\tau}$ be the zeros of the Bessel function $J_{\nu}$ :
\begin{equation}
J_{\tau +3/2}(x_{\xi,\tau})=0,\;\;\xi=1,2,...;\tau=0,1,2,...
\label{zero}
\end{equation}
Then, due to the substitution introduced in Eqs.(2.21) and (2.5) one obtains:
\begin{equation}
E_{\xi,\tau}=\frac{\hbar^2}{2B}k^2_{\xi,\tau},\;\;k_{\xi,\tau}=\frac{x_{\xi,\tau}}{\beta_w}.
\label{Exi}
\end{equation}
Concluding, the differential equation for the beta deformation corresponding to an infinite well
potential provides the energy spectrum given by Eq.(\ref{Exi}) and the wave functions:
\begin{equation}
f_{\xi,\tau}=C_{\xi,\tau}\beta^{-3/2}J_{\tau+3/2}(\frac{x_{\xi,\tau}}{\beta_w}\beta),
\label{fxi}
\end{equation}
where $C_{\xi,\tau}$ is a normalization factor. 

It is worth noticing that the spectra corresponding to E(5) and Davidson potentials become directly comparable by establishing the formal correspondence $n=\xi -1$.


\vskip0.3cm
\subsubsection{A hybrid model }
\vskip0.3cm
In ref.\cite{Rad05} we advanced the idea that the critical point for a phase transition
is depending on  the nuclear state. Therefore the system may reach the critical point in a state of angular momentum J, but in a less excited state, like $(J-2)^+$, the system could behave according to the initial nuclear phase.

According to Ref.\cite{Rad05} the potential energy in the beta variable is depending on angular momentum in the following way:

\begin{eqnarray}
u(\beta)=\left \{\matrix{\beta^2,& \rm{if}&\;\;0\leq \beta <\infty,&\;\;L\leq 2,\cr  
                         0,     & \rm{if}&\;\;0\leq\beta \leq \beta_w,&\;\;L\geq 4,\cr
                         \infty,& \rm{if}&\;\;\beta_w < \beta <\infty,&\;\; L\geq 4.}\right.
\label{redpot}
\end{eqnarray}
The states of interest and their energies have the following expressions:
\begin{eqnarray}
|L^+_{n\tau}M\rangle &=& \sqrt{\frac{2n!}{\Gamma(n+\tau+5/2)}}\beta^\tau
L^{\tau +3/2}_n(\beta^2)e^{-\beta^2/2}G^{LM}_{n\tau}(\gamma,\Omega), \nonumber\\
E_{n\tau} &=& \frac{\hbar^2}{2B}(2n+\tau+5/2),\;\;(n,\tau)=(0,0),\;(0,1),\;\;L=2\tau,\nonumber\\
|L^+_{\xi,\tau}M\rangle &=& C_{\xi,\tau}\beta^{-3/2}J_{\tau+3/2}
(\beta x_{\xi,\tau}/\beta_w)G^{LM}_{\xi-1,\tau}(\gamma,\Omega),\nonumber\\
E_{\xi,\tau} &=& \frac{\hbar^2}{2B}\frac{x_{\xi,\tau}^2}{\beta^2_w},\;\;
(\xi,\tau )=(1,2),\;(1,3),\;(2,0).
\label{states}
\end{eqnarray}
The factor functions depending on the beta variable are solutions of
Eq.(\ref{epsf}) with the reduced potential given by Eq.(\ref{redpot}).
The equation for $\gamma$ deformation and Eulerian angles ($\Omega$)
has the solution $G^{LM}_{n\tau}$.
A possible excited state phase transition was pointed out in Ref.\cite{Rad06}, by using a sixth order solvable boson Hamiltonian. A  potential of an intrinsic deformation radial variable $r$, involving a centrifugal term and a $r^2+r^4$ term, shows up. In the state lying at the top of the barrier separating
the two wells of the potential, the system undergoes a phase transition. This issue has been also addressed by studying the energies and wave functions singularities for Lipkin model as well as for a two level boson model \cite{Caprio07}. 

Note that in all treatments mentioned above, no potential in $\gamma$ is considered. Due to this fact the spectra and wave functions are labeled by the seniority quantum number $\tau$. This feature does not hold when we switch on the $\gamma$-depending potential and moreover impose variable separability by approximating the terms depending on $\gamma$


\subsection{The description of  $\gamma$ degree of freedom}

\noindent
Let us consider the Hamiltonian

\begin{equation}
H=-{\frac{1}{\sin {3\gamma }}}{\frac{\partial }{\partial \gamma }}\sin {
3\gamma }{\frac{\partial }{\partial \gamma }}+U(\gamma)+W(\gamma,Q),  \label{1}
\end{equation}
where $U$ is a periodic function in $\gamma$ with the period equal to $2\pi$ and

\begin{equation}
W(\gamma, Q)={\frac{1}{4}}\sum_{k=1}^{3}{\frac{1}{\sin ^{2}{(\gamma -{\frac{2\pi }{3}}k)%
}}}Q_{k}^{2}  \label{2}
\end{equation}
with $Q_k$ denoting the components of the intrinsic angular momentum.

\subsubsection{Violating some basic properties}
 Any approximation for the potential, by expanding it in power series of $\gamma$, alters the periodic behavior of the eigenfunction. Moreover, the approximating Hamiltonian loses its hermiticity with respect to the scalar product defined with the measure for the gamma variable, $|\sin(3\gamma)|d\gamma$.

We illustrate this by considering the case of a little more complex potential 
\begin{equation}
U=u_1\cos (3\gamma )+u_2\cos ^{2}(3\gamma ).  \label{9}
\end{equation}

Performing the change of function   $\varphi =
\sqrt{\left| \sin (3\gamma )\right| }\psi $, the eigenvalue equation  $H\psi =E\psi $, becomes $\tilde{H}\varphi =0$, with

\begin{equation}
\tilde{H}=\frac{\partial ^{2}}{\partial \gamma ^{2}}+\frac{9}{4}\left[ 1+%
\frac{1}{\sin ^{2}(3\gamma )}\right] -U-W+E.  \label{8}
\end{equation}

We shall consider two situations:

{\it A. Suppose that $\left| \gamma \right| \ll 1$.}    Expanding the terms in $\gamma$ in power series up to the fourth order, one obtains:
\begin{eqnarray}
U_{4}&=&u_1+u_2-9\gamma ^{2}\left( \frac{u_1}{2}+u_2\right) +27\gamma ^{4}\left( 
\frac{u_1}{8}+u_2\right) ,\nonumber  \\
W_{4} &=&\frac{1}{3}\left( 1+2\gamma ^{2}+\frac{26\gamma ^{4}}{9}\right)
\left( Q_{1}^{2}+Q_{2}^{2}\right)\\ 
&&+\frac{2\sqrt{3}\gamma }{9}
 \left(1+2\gamma ^{2}\right) \left( Q_{2}^{2}-Q_{1}^{2}\right)   \label{16}
\nonumber \\
&&+\frac{1}{4}\left( \frac{1}{\gamma ^{2}}+\frac{1}{3}+\frac{\gamma ^{2}}{15}%
+\frac{2\gamma ^{4}}{189}\right) Q_{3}^{2}.  \nonumber
\label{19}
\end{eqnarray}
The low index of U and W suggests that the expansions  in $\gamma$ were truncated at the fourth order. Details about the approximations involved in the following derivation may be found in Appendix A.

Note that due to the term W, the equations of motion for the variable $\gamma$ and and Euler angles are coupled together. Such a coupling term can in principle be handled as we did for the harmonic liquid drop in ref. \cite{Ghe78,Rad78}. 
Here,  we separate the equation for $\gamma$ by averaging $W_4$ with an eigenfunction for the intrinsic angular momentum squared.
 The final result for $H_4$ is:

\begin{eqnarray}
H_{4} 
&=&\frac{\partial ^{2}}{\partial \gamma ^{2}}+\frac{1}{4\gamma ^{2}}\left(
1-\left\langle Q_{3}^{2}\right\rangle \right) +h_{0}+h_{2}\gamma
^{2}+h_{4}\gamma ^{4}\nonumber\\
&&+\frac{2\sqrt{3}\gamma }{9}\left( 1+2\gamma ^{2}\right)
\left\langle Q_{2}^{2}-Q_{1}^{2}\right\rangle ,  \\
h_0&=&E -\frac{1}{3}L(L+1)+\frac{1}{4}\left\langle Q^2_3\right\rangle 
-(u_0+u_1+u_2)+\frac{15}{2},
\nonumber\\
h_2&=&-\frac{2}{3}L(L+1)-\frac{13}{20}\left\langle Q^2_3\right\rangle+
\frac{9}{2}u_1+9u_2+
\frac{27}{20},\nonumber\\
h_4&=&-\frac{26}{27}L(L+1)-\frac{121}{126}\left\langle Q^2_3\right\rangle -\frac{27}{8}u_1-27u_2+\frac{27}{14}\nonumber.
\end{eqnarray}
where L denotes the angular momentum.
If the average is made with the Wigner function $D^L_{MK}$, important simplifications are obtained since the following relations hold:
\begin{equation}
\left\langle Q_{2}^{2}-Q_{1}^{2}\right\rangle =0,\;\;
\left\langle Q_{3}^{2}\right\rangle =K^2
\end{equation}
Actually, this is the situation considered in the present paper.
Note that $H_4$ contains a singular term in $\gamma$, at $\gamma=0$, coming from the term coupling the intrinsic variable $\gamma$ with the Euler angles. One could get rid of such a coupling term by starting with a potential in gamma containing a singular term which cancels the contribution produced by the W term. Thus, the new potential would be

\begin{equation}
U'=U+\frac{9K^2}{4\sin^2(3\gamma)}.
\label{uprime}
\end{equation}
The corresponding fourth order expansion for the Hamiltonian is:

\begin{eqnarray}
H^{\prime}_{4} 
&=&\frac{\partial ^{2}}{\partial \gamma ^{2}}+\frac{1}{4\gamma ^{2}} +h^{\prime}_{0}+h^{\prime}_{2}\gamma
^{2}+h^{\prime}_{4}\gamma ^{4} ,\\
h^{\prime}_0&=&h_0+K^2,\;
h^{\prime}_2=h_2+\frac{27}{20}K^2,\;
h^{\prime}_4 = h_4+\frac{27}{14}K^2\nonumber.
\end{eqnarray}
Some remarks concerning the equation $H^{\prime}_{4}\varphi=0$  are worth to be mentioned: 

i) If in this equation one ignores the $\gamma^4$ term, the resulting equation has the Laguerre functions as solutions and moreover the Hamiltonian
exhibits the $X(5)$  features. 

ii) Note also that the Hamiltonian coefficients are different from those of Ref.\cite{Fortu2}. The difference is caused by the fact that 
here, the expansion is complete.

iii) Taking in the expanded potential $u_1=u_2=0$ and ignoring, for $\gamma$ small, the term $\frac{27}{20}K^2\gamma^2$, the resulting potential is that of an infinite square well which was treated by Iachello in Ref\cite{Iache9}. The solutions are, of course, the Bessel functions of half integer indices. 

iv) Irrespective of the potential in $\gamma$, in the regime of $|\gamma|$ small a term proportional to $\gamma^2$ shows up due to the rotational Hamiltonian $W$. Therefore, even in the case the potential is taken as an infinite square well, of the form $1/\gamma^2$, the equation describing the $\gamma$ variable admits a Laguerre function as solution and not, as might be expected, a Bessel function of semi-integer index.
Amazingly, the potential in $\gamma$ is also of Davidson type.

v) None of the mentioned solutions is periodic. 

vi) Also  the approximated Hamiltonians are not Hermitian in the Hilbert space of functions in gamma with the integration measure as introduced by the liquid drop model, i.e. $|\sin3\gamma|d\gamma$.

{\it B. The case $|\gamma-\pi/6|\ll 1$.}
Using the fourth order expansion in $y=|\gamma-\pi/6|$, given in Appendix A,
one obtains a Hamiltonian similar to that given by Eq. (2.36):
\begin{equation}
H^{\prime}_4=\frac{\partial}{\partial \gamma^2}+h^{\prime}_2\gamma^2+h'_4\gamma^4+2\sqrt{3}y\left(1+\frac{22\sqrt{3}}{3}y^2\right)\langle Q_3^2-Q_2^2\rangle.
\end{equation}
If $\langle Q_3^2-Q_2^2\rangle =0$ and, moreover, one ignores the term in $\gamma^4$ the resulting equation in $\gamma$ describes a harmonic oscillator. Again the eigenfunctions, i.e. the Hermite functions, are orthogonal on an unbound interval of $\gamma$, and not on $[0,2\pi]$.

\subsubsection{Toward an exact treatment which preserves periodicity and hermiticity}

In order to remove the drawbacks mentioned above,  we try first to avoid making approximations. Thus, let us consider the Hamiltonian given by Eq.(\ref{1}) where instead of U we consider  $ U^{\prime}$ as defined by Eq.(\ref{uprime}), and ignore for a moment $W$.  Changing the  variable $x=\cos3\gamma$,  the eigenvalue equation associated to this Hamiltonian, $HS=ES$, becomes:

\begin{equation}
 \left( 1-x^{2}\right) \frac{{\rm d}^{2}S}{{\rm d}\,x^{2}}-
2x\frac{{\rm d} S}{{\rm d}\,x}+
\left(\frac{1}{9}(E-u_{1}x-u_{2}x^{2})-\frac{K^{2}}{4{(1-}x^{2}{)}}\right) S=0.
\label{99}
\end{equation}
Note that we denoted the eigenfunction by $S$ which suggests that the differential equation
(\ref{99}) is obeyed by a spheroidal function.
If $u_1=u_2=K=0$, the solution of this equation is the Legendre polynomial $P_n$ while $E=9n(n+1)$. This case has been considered in Ref.\cite{Fortu2}. This function may be used  to approximate the solution of the original liquid drop model. 
For other particular choices of the coefficients $u_1,u_2$ defining the potential in gamma, the solution is readily obtained if one compares the above equation with that characterizing the spheroidal oblate functions \cite{Abra}
\begin{eqnarray}
\left( 1-x^{2}\right) \frac{{\rm d}^{2}S_{nm}}{{\rm d}\,x^{2}}-2x\frac{{\rm d}S_{nm}}
{{\rm d}\,x}+\left(\lambda
_{nm}-c^{2}x^{2}-\frac{m^{2}}{{1-}x^{2}}\right) S_{nm}=0.  \label{10}
\end{eqnarray}
The prolate case is reached by changing $c\to {\rm i}c$.

 For $c=0$, the solutions of  Eq.(\ref{10}) are the associated Legendre functions
 $P_{n}^{m}$. For $c\ne 0$, $S_{nm}$, with $m,n$ integers and $n \ge m\ge 0$,
are linear series of these functions.

In the case $u_{1}=0$, the solution of Eq.(\ref{99}) is identified as being the spheroidal function while  the energy is simply related to $\lambda_{nm}$: 
\begin{equation}
m=\frac{K}{2},\quad c^{2}=\frac{u_{2}}{9},\quad \lambda _{nm}=\frac{1}{9}%
E_{nm}.  \label{11}
\end{equation}
Here $E_{nm}$ denotes the eigenvalue $E$ corresponding to the quantum numbers $n$ and $m$.

\begin{figure}[h!]
\begin{center}
\includegraphics[height=8cm]{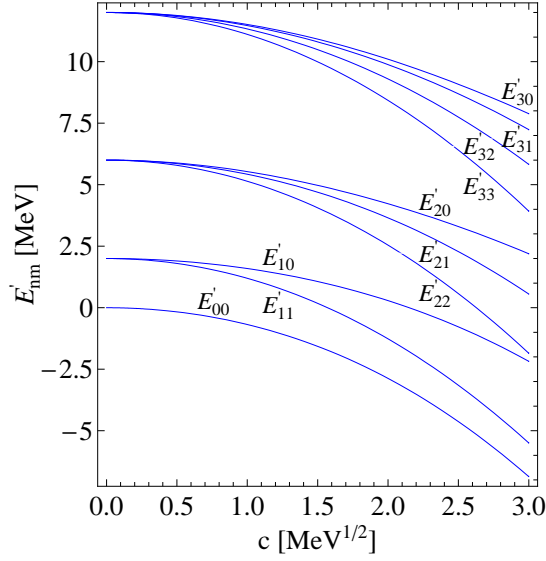}
\end{center}
\caption{(Color online).The spheroidal energy $E^{\prime}_{nm}=\lambda_{nm}=E_{nm}/9$, for $0\le m\le n\le 3$  are plotted as functions of 
$c=\sqrt{u_2}/3 $.}
\label{Fig. 1}
\end{figure}

For $|c|$ small the energies $E_{nm}$ exhibits the asymptotic expansion
\begin{eqnarray}
&&E_{nm}\approx 9n(n+1)-\frac{2\left(n(n+1)+m^2-1\right)}{(2n-1)(2n+3)}u_2
\label{12}\\
&+&\frac{1}{18}\frac{\left[(n-1)^2-m^2\right](n^2-m^2)}{(2n-3)(2n-1)^3(2n+1)}
u^2_2\nonumber\\
&-&\frac{1}{18}\frac{\left[(n+1)^2-m^2\right]\left[(n+2)^2-m^2\right]}
{(2n+1)(2n+3)^3(2n+5)}u^2_2.
\nonumber
\end{eqnarray}
 Eq.(\ref{12}) considered for a fixed $m$ but various $n$, defines a band. Similar expansions may be derived for $|c|$ large.

\begin{eqnarray}
E_{nm}&\approx& -u_2+3q\sqrt{u_2}+9\left(m^2-\frac{q^2+5}{8}\right)-\frac{27q}{64\sqrt{u_2}}(11+q^2-32m^2),\nonumber\\
q&=&2(n-m)+1
\end{eqnarray}
We remark that the spectrum has a rotational behavior for small $c$, due to the term $n(n+1)$ while for large values of $c$ it has an oscillator feature, the energy depending linearly on $n$. 

If one needs the expansion up to the $1/c^2$ terms, the results for the first few energies are:
\begin{eqnarray}
E_{11}&=&9\left(\frac{1}{4}-c^2+c+\frac{5}{16c}+\frac{33}{64c^2}\right),\nonumber\\
E_{21}&=&9\left(-\frac{3}{4}-c^2+3c+\frac{9}{16c}+\frac{135}{64c^2}\right),\nonumber\\ 
E_{22}&=&9\left(\frac{13}{4}-c^2+c+\frac{29}{16c}+\frac{177}{64c^2}\right),\nonumber\\
E_{31}&=&9\left(-\frac{11}{4}-c^2+5c-\frac{5}{16c}+\frac{219}{64c^2}\right),\nonumber\\
E_{32}&=&9\left(\frac{9}{4}-c^2+3c+\frac{81}{16c}+\frac{855}{64c^2}\right),\nonumber\\
E_{33}&=&9\left(\frac{33}{4}-c^2+c+\frac{69}{16c}+\frac{417}{64c^2}\right).
\end{eqnarray}

It is worth spending few words about Fig. 1 where the energies correspond to the spheroidal functions with the parameters specified by Eq.(2.40). Indeed for $c\to0$ one notices some multiplet degeneracy which suggest a symmetry with respect to K, i.e. a rotation invariance of states of a given $n$. Increasing $c$ the split in energy is similar to that in Nilsson \cite{Nils} model when the energy is $\Omega$ dependent. The difference is that while in Nilsson model each deformed state is a superposition of states with different angular momentum, here the multiplet members are characterized by the same $n$. In this respect the feature shown in Fig.1 is similar to the one obtained with a spherical projected single particle basis
\cite{RadIud}. In the region of large $c$, for a given large $n$ the set of states of different $m$ seem to form a band. On the other hand for a fixed $m$ the set of states with different n is a band of equidistant energy levels. 

\subsection{Approximation which does not affect periodicity and hermiticity}
Now, we shall focus on an approximate solution which preserves the periodicity in $\gamma$. For that purpose we consider the Hamiltonian
\begin{eqnarray}
H&=&-\frac{1}{\sin 3\gamma}\frac{\partial}{\partial \gamma}\sin 3\gamma \frac{\partial}{\partial \gamma}+U(\gamma),\nonumber\\
U(\gamma) &=&u_1\cos3\gamma+u_2\cos^23\gamma+\frac{K^2}{4\sin^2\gamma}.
\end{eqnarray}
Changing the function by the transformation 
$\Psi=|\sin(3\gamma)|^{-1/2}{\Phi}$, for $\sin(3\gamma )\ne 0$, the eigenvalue equation for H is  $\tilde{H}\Phi =0$ with $\tilde{H}$ given by Eq. (2.31) for $W=0$.

Under the regime of $|\gamma|$ small, we take the $O(\gamma ^3)$ expansion of the terms depending on $\gamma$ and in the final expression  approximate $\gamma \approx \sin\gamma$. In this way the eigenvalue equation becomes:

\begin{eqnarray}
&&\left(\frac{\partial^2}{\partial \gamma^2}+a-2q\cos2\gamma-\frac{K^2-1}{4\sin^2\gamma}\right){\Phi}=0,\;\rm{with}\label{26}\\
&&q=\frac{1}{3}+\frac{9}{8}u_1+\frac{9}{4}u_2,u=u_2+\frac{347}{108},a=E+\frac{10}{9}q+u\nonumber .
\end{eqnarray}
We suppose now that this equation is valid in the interval $[0,2\pi]$.
The equation (\ref{26}) is just the trigonometric form of the spheroidal functions. The algebraic version is obtained  by changing the variable $x=\cos\gamma.$

For $K=1$ one obtains the Mathieu equation:
\begin{equation}
\left(\frac{\partial^2}{\partial \gamma^2}+a-2q\cos2\gamma\right)\Phi=0.
\end{equation}
\noindent
There are two sets of solutions, one even and one odd denoted by $\Phi^+(a,q,\gamma)$ and $\Phi^-(a,q,\gamma)$, respectively. For $q=0$, both solutions are periodic for any positive value of $a$.
\begin{equation}
\Phi^+(a,0,\gamma)=\cos\left(\sqrt{a}\gamma\right),\;\Phi^-(a,0,\gamma)=
\sin\left(\sqrt{a}\gamma\right).
\end{equation}

\begin{figure}[h!]
\begin{center}
\includegraphics[height=6cm]{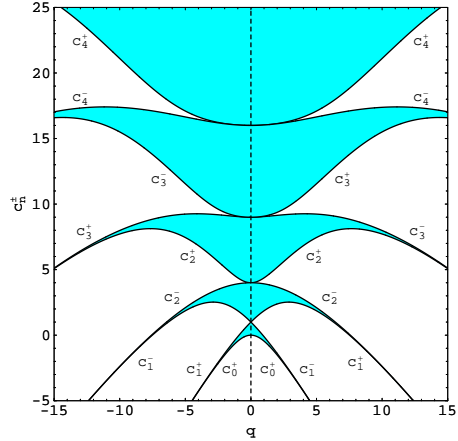}
\end{center}
\caption{(Color online)The characteristic curves $c^{\pm}_n$  are plotted as functions of q for several values of n.}
\label{Fig. 2}
\end{figure}

For $q\ne 0$ the Mathieu functions are periodic in $\gamma$ only for a certain set of values of $a$, called characteristic values. These are denoted by $c^+_n$ for even and $c^-_n$ for odd functions, respectively. In the plane $(a,q)$, the characteristics curves $c^{\pm}_n$   separate the stability regions, shown in Fig. 2 by gray color, from the non-stability  ones, indicated by white color in the quoted figure.
For $q=0$ the equalities $c^{\pm}_n(0)=n^2$ hold.
By means of Eq.(\ref{26}) the characteristic values determine the energy E.
Thus, the energy spectrum is given by $E^{\pm}_n-u$ with $E^{\pm}_n=c^{\pm}_n-\frac{10}{9}q$. The corresponding wave functions are the elliptic cosine and elliptic sine functions respectively:
\begin{eqnarray}
 \Phi^+_0(q,\gamma)&=&\frac{1}{\sqrt{2\pi}}ce_0(q,\gamma),
\Phi^+_n(q,\gamma)=\frac{1}{\sqrt{\pi}}ce_n(q,\gamma),\nonumber\\
\Phi^-_n(q,\gamma)&=&\frac{1}{\sqrt{\pi}}se_n(q,\gamma),\;n=1,2,...
\end{eqnarray}

They form an orthogonal set. The matrix elements of the gamma depending factors of the transition operator can be easily calculated in Mathematica. Moreover, in the regime of $|q|$-small these matrix elements can be analytically performed, since the following representation of the wave functions hold:
\begin{equation}
\Phi^{\pm}_n(\gamma)\approx \cos(n\gamma-\theta_{\pm})-
\left[\frac{\cos\left[(n+2)\gamma-\theta_{\pm}\right]}{4(n+1)}-
\frac{\cos\left[(n-2)\gamma-\theta_{\pm}\right]}{4(n-1)}\right]q^2.
\end{equation}
where $n\ge 3, \theta_{+}=0$ and $\theta_-=\pi/2$. The corresponding energies have very simple expressions:
\begin{eqnarray}
E^+_0 &\approx &u-\frac{10}{9}q-\frac{q^2}{2},\nonumber\\
E^+_1 &\approx &E^-_1 \approx u-\frac{10}{9}q-\frac{q^2}{8},\nonumber\\
E^+_2 &\approx &E^-_2 \approx u+4-\frac{10}{9}q-\frac{q^2}{2},\nonumber\\
E^+_n &\approx &E^-_n \approx u+n^2-\frac{10}{9}q-\frac{q^2}{2(n^2-1)},\;n\ge 3.
\end{eqnarray}

Normalizing the above functions to unity, on the $\gamma$ interval $[-\pi,\pi]$ with respect to the integration measure $d\gamma$ and calculating with the resulting functions the matrix elements of the $\gamma $  depending factors involved in the electric transition operator one obtains the curves represented in Fig.3.

\begin{figure}[h!]
\begin{center}
\includegraphics[height=6cm]{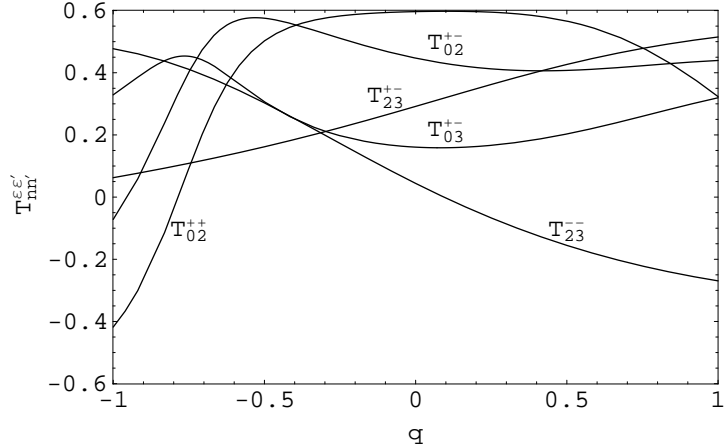}
\end{center}
\caption{The matrix elements of  $T^{\epsilon\epsilon'}_{nn'}$ for $\cos\gamma$ ($\epsilon\epsilon'=+$) and $\sin\gamma$ ($\epsilon\epsilon'=-$) are represented as functions of $q$.}
\label{Fig. 3}
\end{figure}

Obviously, a phase transition is determined by the combined effects coming from
the behavior of the wave function in the $\beta$ and $\gamma$ variables, respectively.

{\it In the $X(5)$ formalism, the eigenfunction of the $\beta$ Hamiltonian is a Bessel function of irrational index, while the $\gamma$ Hamiltonian's eigenfunction is a Laguerre polynomial.}

 Here we propose to change the description in the $\gamma$ space either by a spheroidal or by a Mathieu function. These functions are periodic and the corresponding Hamiltonians Hermitian.
Moreover, in both versions, the $X(5)$ Hamiltonian is obtained in the limit of small $|\gamma |$.
In Ref.\cite{Baerd} a periodic $\gamma$ potential with a minimum in $\gamma_0=\pi/6$, was considered. The model is solvable and the wave function is a Legendre polynomial. Moreover,  energies are analytically obtained. By contrast the situation considered here is more complex, the energies being obtained by solving the eigenvalue equation for the spheroidal functions. However the picture described in Ref.\cite{Baerd} is recovered under some particular restrictions.

\subsection{Including the rotational term preserves periodicity and hermiticity}

We recall that so far the rotational term $W$ was left out. Now we turn our attention to this term. If we average $W$ with the Wigner function $D^L_{MK}$ and add the result to the potential $U'$ given by Eq.(2.35) and then following the same path as before, one ends up also with an equation for a spheroidal function. Indeed, let us consider the average of W:
\begin{eqnarray}
\langle LK|W|LK\rangle &=&\frac{9D}{8\sin^23\gamma}-\frac{D-2K^2+2}{8\sin^2\gamma},\nonumber\\
D&=&L(L+1)-K^2-2.
\end{eqnarray}
When  $|\gamma|\ll 1$, this expression admits the following second order expansion in $\sin\gamma$:
\begin{equation}
W(\gamma)=\frac{K^2-1}{4\sin^2\gamma}+\frac{1}{3}\left[L(L+1)-K^2-2\right]\left(1+2\sin^2\gamma \right).
\end{equation}
The term $L(L+1)/3$ from the above expression, multiplied with the factor $1/\beta^2$ are added to the equation describing the variable $\beta$. In the case one makes the option for a infinite well potential in $\beta$, the renormalization just mentioned leads to an equation in 
$\beta$, whose solution is a Bessel function with the index 
\begin{equation}
\nu=\left(\frac{1}{3}L(L+1)+\frac{9}{4}\right)^{1/2}
\end{equation}

For the case $|\gamma|\ll 1$, we consider the second order expansion in $\sin\gamma$ for the full Hamiltonian (2.28). The result is a trigonometric form for the equation of the spheroidal function:
\begin{equation}
\frac{\partial ^2\varphi}{\partial \gamma^2}+\left(E'-\frac{K^2-1}{4\sin^2\gamma}-C\sin^2\gamma\right)\varphi =0,
\label{sphtrig}
\end{equation}
with the notations:
\begin{equation}
E'=E+\frac{9}{4}-u_1-u_2-\frac{D}{3},~~C=\frac{2D}{3}-\frac{9u_1}{2}-9u_2.
\label{enEprim}
\end{equation}

As suggested by the expression of the starting Hamiltonian, the remaining terms of $W(\gamma)$  should be multiplied with $1/\beta^2$. This coupling of 
$\beta$ and $\gamma$ variables is usually considered as a renormalization term for the potential in $\gamma$, by replacing the factor $1/\beta^2$ by the constant $1/\langle\beta^2\rangle$. The notation $\langle \beta^2\rangle$ is used for the expectation value of $\beta^2$ in the ground state which results in redefining the constant $C$, in Eq. (2.55).

Eq. (\ref{sphtrig}) can be brought to the form given by Eq.(2.39) by making a successive change of function
$S=\left|\sin \gamma|\right)^{-1/2}\varphi$ and variable, $x=\cos\gamma$. Indeed, the resulting equation is
that of the spheroidal function defined by:
\begin{eqnarray}
\lambda_{nm}&=&E_{nm}+\frac{7}{2}u_1+8u_2+2-D+\frac{1}{3}L(L+1),\nonumber\\
c^2 &= &\frac{9}{2}u_1+9u_2-\frac{2}{3}D.
\end{eqnarray}
This equation has been used by some authors of this paper, in Ref.\cite{Rad07}, to describe the spectrum and the E2 properties of $^{152}$Sm.
 \subsection{Recovering X(5) in the limit of $|\gamma|$-small}
It is worth comparing the present formalism based on the spheroidal functions with the X(5) approach. It is easy to prove that, indeed, X(5) is the limiting case for our approach. Indeed, considering the second order expansion in $\gamma$ of the terms involved in Eq.(\ref{sphtrig}) one arrives at:
\begin{equation}
\frac{\partial^2\varphi}{\partial\gamma^2}+\left[E'-\frac{K^2-1}{4}\left(\frac{1}{\gamma^2}+\frac{1}{3}\right)-\left(\frac{K^2-1}{60}+C\right)\gamma^2\right]\varphi=0.
\end{equation}
This equation is characterizing the $X(5)$ model, with all harmonic contributions included. Indeed, changing the variable
$\xi=q\gamma$, with $q=\sqrt{C+\frac{K^2-1}{60}}$, the differential equation becomes:
\begin{equation}
\frac{d^2\varphi}{d\xi^2}+\left[\frac{1}{q^2}\left(E'+\frac{1-K^2}{12}\right)-\xi^2-\left(\alpha^2-\frac{1}{4}\right)\frac{1}{\xi^2}\right]\varphi ,~\alpha=\frac{K}{2}.
\end{equation}
Comparing this equation with that describing an harmonic, isotropic plane oscillator:
\begin{equation}
\frac{d^2\Phi_{n\alpha}}{d\xi^2}+\left[2(n+\alpha+1)-\xi^2-\left(\alpha^2-\frac{1}{4}\right)\frac{1}{\xi^2}\right]
\Phi_{n\alpha}=0 ,~~\alpha=\frac{K}{2}.
\end{equation}
one identifies the function $\varphi $ with the function Laguerre:
\begin{equation}
\varphi_{n\alpha}=\sqrt{2\frac{n!}{\Gamma(n+\alpha+1)}}\xi^{\alpha}L^{\alpha}_n(\xi^2)\exp\left(-\frac{1}{2}\xi^2\right),
\end{equation} 
while the system energy is:
\begin{equation}
E_n=2\left(C+\frac{K^2-1}{60}\right)^2\left(n+\frac{K}{2}+1\right)+u_1+u_2+\frac{1}{4}K^2-3.
\end{equation}
The property of reaching the $X(5)$ model in the limit of small values of $|\gamma|$, holds also for the Mathieu functions. Indeed, these functions satisfy a differential equation which is of spheroidal type. Consequently, in the limit $|\gamma|\ll 1$, the Mathieu functions may account for the properties which are specific to the $X(5)$ approach.
Numerical applications with Mathieu functions will be published elsewhere.

\renewcommand{\theequation}{3.\arabic{equation}}
\setcounter{equation}{0}
\section{The present approach}
\label{sec:level3}

Here we summarize the procedure adopted in the present paper, to treat a phenomenological solvable Hamiltonian defined in the space of the variables $\beta$ and $\gamma$.

The potential in the two variables is considered to be of the form given by Eq. (2.3). As $V(\beta)$ we take the Davidson potential (2.14). Including  the terms proportional to $\frac{1}{\beta^2}$ (this is $\frac{1}{3\beta^2}\left[L(L+1)\right]$) from the rotational term  in the Schr\"{o}dinger equation for $\beta$ , the  resulting equation admits solutions which formally coincide with those given by Eqs. (2.8) and (2.9) but having instead of  $\tau$ a irrational quantum number $p$ defined as:
\begin{equation}
p=-\frac{3}{2}+\left[\frac{1}{3}L(L+1)+\frac{9}{4}+\beta_0^4\right]^{1/2}
\end{equation}
This expression is obtained by writing the coefficient of $\frac{1}{\beta^2}$ from the Schr\"{o}dinger equation associated to the $\beta$ variable in the form:
\begin{equation}
(p+1)(p+2)=2+\frac{1}{3}L(L+1)+\beta_0^4.
\end{equation}
This equation has two solutions, one written above, while the second one is differing from the first one by the sign of the square root term. Let us denote for a while the two solutions by $p_{\pm}$. Note that the Davidson potential is not a continuous function in $\beta =0$. This causes the fact that for L=0, we have: 
\begin{equation}
\lim_{\beta_0\to 0}p_+=0,~~ \lim_{\beta_0\to 0}p_-=-3.
\end{equation}
Consequently, the corresponding spectra are given by:
\begin{eqnarray}
E^{(+)}_n&=&2n+\frac{5}{2},\nonumber\\
E^{(-)}_n&=&2n-3+\frac{5}{2}.
\end{eqnarray}
Therefore, the full spectrum of the 5-dimensional oscillator is recovered only if both solutions are considered at a time
\cite{Palma}.
Since, as we shall see a bit later,  $\beta_0$ is far from origin we make the option for the branch corresponding to the eigenvalue $p_+$. The choice is justified by the fact that  for $\beta_0\ne 0$ the wave function corresponding to $p_-$ is singular in origin and therefore it is not a convenient solution.

As explained before, for the states belonging to the ground band  $\beta_0$ was fixed variationally, by Eq.(2.17). We extended Eq. (2.17) to the beta and gamma bands, respectively. The first derivatives of the ratios $R^{(k)}_L$, defined by
\begin{eqnarray}
R^{(\beta)}_L&=&\frac{E_{L^+_{\beta}}-E_{0^+_{\beta}}}{E_{2^+_{\beta}}-E_{0^+_{\beta}}},L\ge 4,\nonumber\\
R^{(\gamma)}_L&=&\frac{E_{L^+_{\gamma}}-E_{2^+_{\gamma}}}{E_{3^+_{\gamma}}-E_{2^+_{\gamma}}},L\ge 4,
\end{eqnarray}
have the $\beta_0$ dependence shown in Fig.4 for some particular values of  $L$.
We fix $\beta_0$ for the states in the band $k$ (=$\beta$, $\gamma$) so that the first derivatives of the ratios
$R^{(k)}_L$ are maximum. From Fig. 4 one sees that each curve has a well pronounced maximum. Collecting the values of $\beta_0$ obtained in this way, and representing them as function of $L$
one obtains a straight line for both beta and gamma bands, as shown in Fig.5. Extrapolating these straight lines for $L=0, 2$ in beta band and $L=2, 3$ in the gamma band, one obtains a one to one correspondence between the states in the two bands and the values of $\beta_0$.   

\begin{figure}[hb!]
\begin{minipage}[t]{8cm}
\epsfysize=5cm
\centerline{\epsfbox{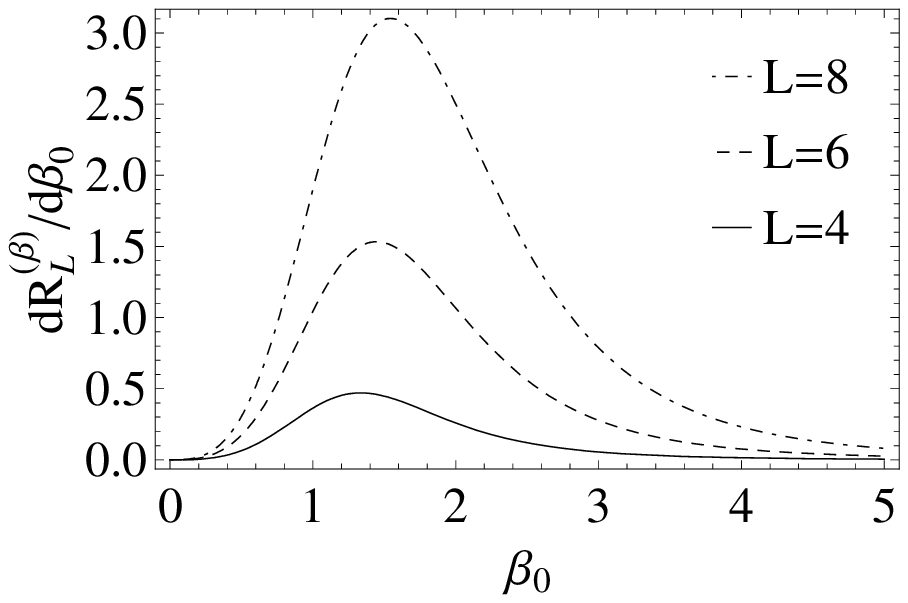}}
\epsfysize=5cm
\centerline{\epsfbox{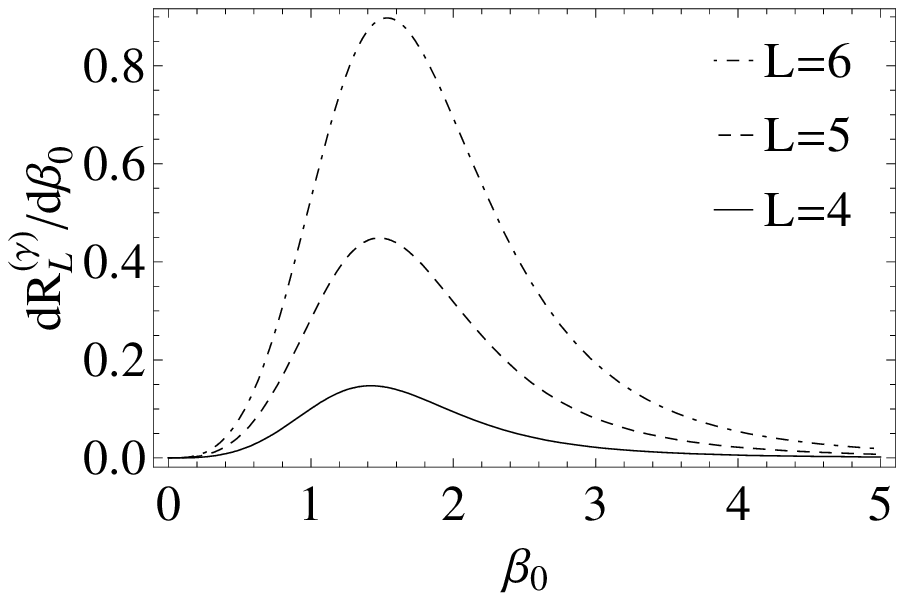}}
\caption{ The first derivative of the ratios $R^{(\beta)}_L$ (upper panel) and $R^{(\gamma)}_L$ (lower panel), defined by Eq. (2.5), are plotted as function of $\beta_0$. }
\end{minipage}
\hspace{\fill}
\begin{minipage}[t]{8cm}
\epsfysize=5cm
\centerline{\epsfbox{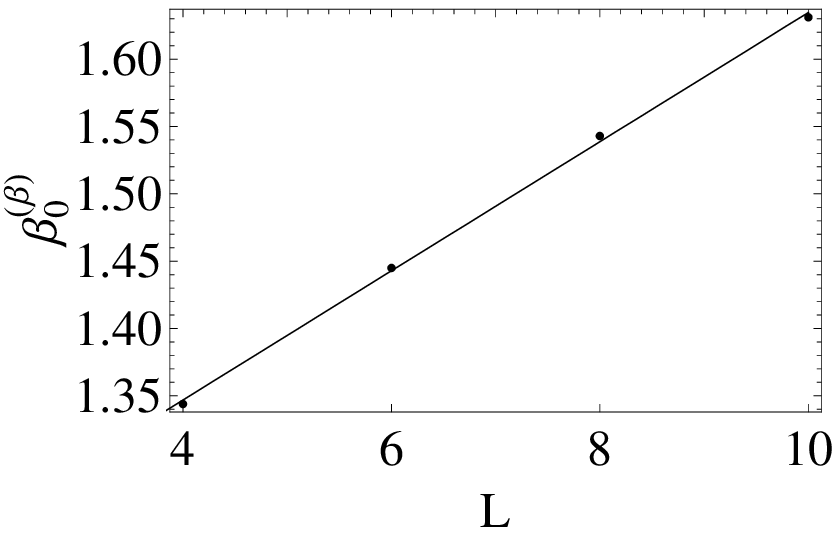}}
\epsfysize=5cm
\centerline{\epsfbox{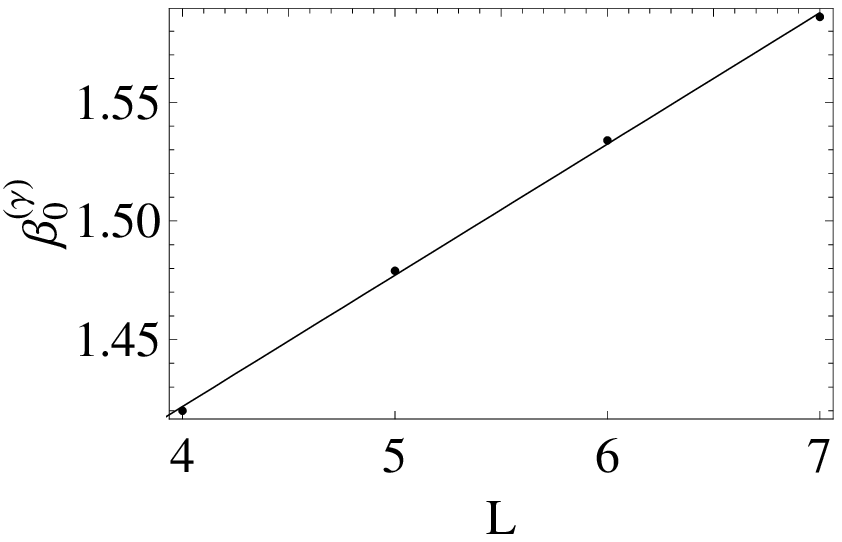}}
\caption{The  solutions $\beta^{(k)}_L$ of the equation $\frac{d^2R^{(k)}}{d\beta_0^2}=0$ for $k=\beta$ (upper panel) and $k=\gamma$ (lower panel), are plotted as functions of the angular momentum $L$. }
\end{minipage}
\end{figure}

\clearpage

Let us denote by $E^{(\beta, D)}_{np} (=2n+p+\frac{5}{2})$ the energy provided by the Shr\"{o}dinger equation associated to the variable $\beta$.
The index D suggests that the $\beta$ potential is chosen to be of Davidson's type.

For comparison we considered also an infinite square well potential for the $\beta$ variable.
In this case the  energies associated to the $\beta$ variable are denoted by $E^{(\beta,B)}_{\xi s}$. They correspond to the Bessel function of index $s+3/2$ with $s$ defined by the following equation:
\begin{equation}
\left(s+\frac{3}{2}\right)^2=\frac{9}{4}+\frac{1}{3}L(L+1)
\label{squn}
\end{equation}
$\xi$ is an ordering index for the zeros of the Bessel function. Therefore the irrational index for the Bessel function will be:
\begin{equation}
s=-\frac{3}{2}+\sqrt{\frac{1}{3}L(L+1)+\frac{9}{4}}.
\end{equation}
Other possible ways of renormalizing the differential equation for $\beta$ are discussed in Appendix C.
The cases when the spheroidal function formalism for the $\gamma$ variable is considered at a time with an oscillator potential or a hybrid potential potential in $\beta$ will discussed elsewhere.    

As for the potential in $\gamma$ we considered
\begin{equation}
U(\gamma)=\frac{1}{\langle \beta^2\rangle}\left[u_1\cos3\gamma+u_2\cos^23\gamma+\frac{9}{4\sin^23\gamma}\right].
\end{equation}
Assuming that $|\gamma|\ll 1$ the rotational term is expanded in powers of $\sin3\gamma$.
From the rotational term we depict the term not depending on $\gamma$ and proportional to $L(L+1)$, otherwise being proportional to $\frac{1}{\beta^2}$, and
add it to the Hamiltonian in $\beta$ which results in having a renormalization of the centrifugal term. In the remaining terms, we approximate $1/\beta^2$ by $1/\langle\beta^2\rangle$. Thus, the Hamiltonian in $\gamma$ will comprise an overall factor 
$1/\langle\beta^2\rangle$. If in the Hamiltonian which multiplies this factor, one changes  the function
$\varphi\to S=|\sin3\gamma|^{-{1/2}}\varphi$ and the variable $\gamma\to x=\cos3\gamma$, the corresponding Schr\"{o}dinger equation is that of a spheroidal function  defined by Eq. (2.39) with

\begin{eqnarray}
\lambda_{nm}&=&\frac{1}{9}\left(E^{(\gamma)}_{nm}-\frac{1}{2}u_1-\frac{11}{27}D+\frac{1}{3}L(L+1)\right),\nonumber\\
c^2&=&\frac{1}{9}\left(\frac{1}{2}u_1+u_2-\frac{2}{27}D\right),\nonumber\\
m&=&\frac{K}{2}.
\end{eqnarray}

For illustration, in Figs. 6, 7 we give few potentials corresponding to $\langle\beta\rangle=1$ and different sets of 
$(u_1,u_2)$. The spheroidal functions normalized to unity in the interval $[0,\pi/3]$, given by Eq. (2.39) with the parameter $c^2$ determined by $(u_1,u_2)$ used in Figs. 6, 7 are represented as functions of $\gamma$ in Figs. 8, 9 for three  
$L$ levels from the ground band. 
Once we fix the $\gamma$  potential, we can calculate the energy associated to the $\gamma$ variable.
 
The total energy for the  system described by the decoupled $\beta$ and $\gamma$ variables is:
\begin{equation}
E^{(k)}_{n\tau;n'm;LK}=E_{0}+AE^{(\beta,k)}_{n\tau}+FE^{(\gamma)}_{n'm}, ~~k=D,B.
\end{equation}  
The coefficient $F$ includes the factor $1/\langle \beta^2\rangle$ mentioned above. Due to this feature there is no need to specify the average value of $\beta^2$.
Note that the energy determined by the rotational degrees of freedom has been already included when the term of the Hamiltonian denoted by $W$ was averaged  with the Wigner functions.

If in the space of $\beta$, the  Schr\"{o}dinger equation involves the Davidson's potential, the wave function describing the whole system is:
\begin{equation}
|np;n'm;LMK\rangle=\Psi_{np}(\beta)S_{n'm}(\cos3\gamma)\frac{\sqrt{2L+1}}{4\pi}\left(D^L_{MK}+(-1)^{L+K}D^L_{M,-K}\right),m=\frac{K}{2}.
\end{equation}
The ground, beta and gamma bands are defined by the quantum numbers:
\begin{eqnarray}
n&=&0, ~ n'=0, ~m=0, K=0, L=0,2,...~~\rm{ground~~band},\nonumber\\
n&=&0,~ n'=1, ~m=1, K=2, L=2,3,...~~\rm{gamma~~band},\nonumber\\
n&=&1,~ n'=0, ~m=0, K=0, L=0,2,...~~\rm{beta ~~band}.
\end{eqnarray}

In the situation when the $\beta$ potential is an infinite square well, the wave function has the expression:

\begin{equation}
|n s;n'm\rangle=f_{\xi s}(\beta)S_{n'm}(\cos3\gamma)\frac{\sqrt{2L+1}}{4\pi}\left(D^L_{MK}+(-1)^{L+K}D^L_{M,-K}\right),m=\frac{K}{2}.
\end{equation}
with $f_{\xi\tau}$ given by Eq.(2.10) and the irrational index $s$ given by Eq.(\ref{squn}) and $S_{n'm}(\cos3\gamma)$ defined by Eq.(2.39).
The quantum numbers defining the ground, beta and gamma bands are as follows:
\begin{eqnarray}
\xi&=&1,~ n'=0, ~m=0, K=0, L=0,2,...~~\rm{ground~~band},\nonumber\\
\xi&=&1,~, n'=1, ~m=1, K=2, L=2,3,...~~\rm{gamma~~band},\nonumber\\
\xi&=&2,~, n'=0, ~m=0, K=0, L=0,2,...~~\rm{beta ~~band}.
\end{eqnarray}.

Once the wave functions are determined by solving the corresponding eigenvalue equations,
 we can proceed to calculating the electric transition probabilities. In order to get  a feeling about how sensitive  the matrix elements of 
$\gamma$ depending terms of the transition operator are to changing the parameter $c$, we have plotted them in Figs.10, 11,  versus 
$c$. From there one notices that in a large interval of $c$, the matrix elements are slowly varying with $c$. The diagonal matrix elements of $\cos\gamma$ (the first panel in the left column) are slightly increasing  by 0.01 starting with the values 0.823, 0.844 and 0.854 at $c=0$. The corresponding matrix elements of $\sin\gamma$ are changing just a little when we vary $c$, starting with the values 0.475 0.487 and 0.493. The magnitudes of the matrix elements between states belonging to the same multiplet (see Fig.1) are small. The matrix elements characterized by the same $\Delta n=1$ are relatively large for $\cos\gamma$ but small for $\sin\gamma$.

The reduced E2 transition probabilities have been calculated by using alternatively  a harmonic, $T^{(h)}$, and an anharmonic transition operator, $T^{(anh)}_{2\mu}$,  having the expressions:
\begin{eqnarray}
T^{(h)}_{2\mu}&=&t\beta \left(\cos\gamma D^2_{\mu 0}+\frac{\sin\gamma}{\sqrt{2}}(D^2_{\mu2}+D^2_{\mu, -2})\right),\nonumber\\  
T^{(anh)}_{2\mu}&=&t_1\beta \left(\cos\gamma D^2_{\mu 0}+\frac{\sin\gamma}{\sqrt{2}}(D^2_{\mu 2}+D^2_{\mu, -2})\right)+\nonumber\\
         & &t_2\sqrt{\frac{2}{7}}\beta^2 \left(-\cos2\gamma D^2_{\mu 0}+\frac{\sin2\gamma}{\sqrt{2}}(D^2_{\mu 2}+D^2_{\mu, -2})\right)
\end{eqnarray}	 
The strengths $t$, $t_1$ and $t_2$ are free parameters which are fixed by fitting one and two particular $B(E2)$ values, respectively. Due to the structure of the wave functions specified above, the matrix elements between the states involved in a given transition are factorized into matrix elements of the transition operators factors depending on $\beta$, $\gamma$ and the Euler angles, respectively.
\begin{figure}[ht!]
\begin{minipage}[t]{8cm}
\epsfysize=6cm
\centerline{\epsfbox{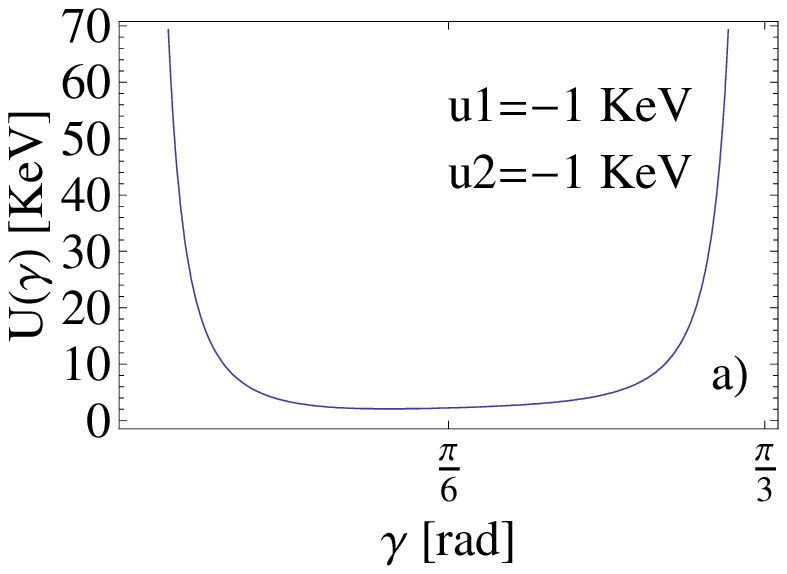}}
\epsfysize=6cm
\centerline{\epsfbox{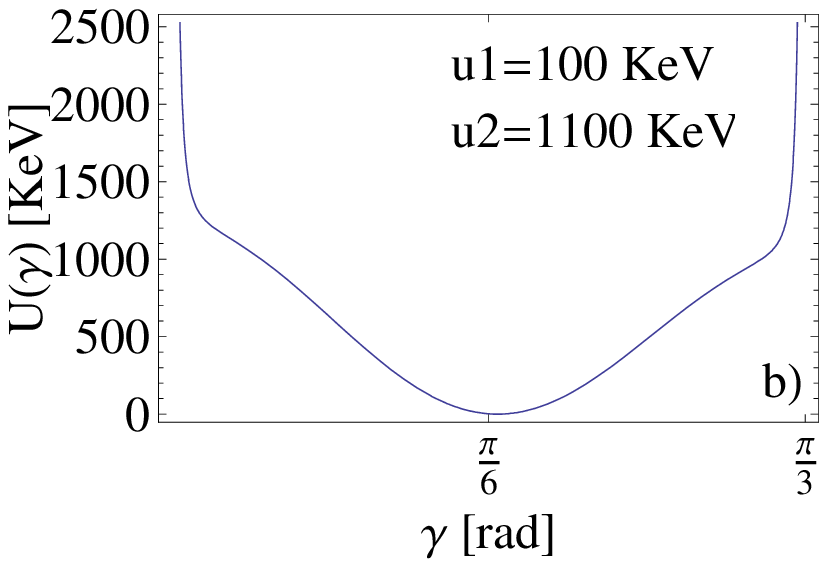}}
\epsfysize=6cm
\centerline{\epsfbox{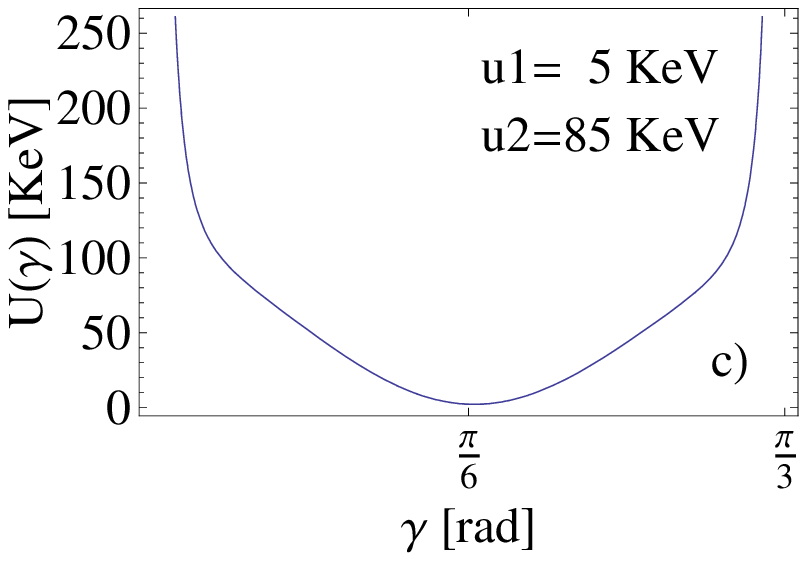}}
\caption{  For three sets of parameters $(u_1,u_2)=(-1,-1);(100,1100);(5, 85) keV$, the $\gamma$ potential $U(\gamma)$ (3.8), with $\langle \beta^2\rangle =1$, is plotted as a function of $\gamma$ in the pannels a), b), c), respectively. }
\end{minipage}
\hspace{\fill}
\begin{minipage}[t]{8cm}
\epsfysize=6cm
\centerline{\epsfbox{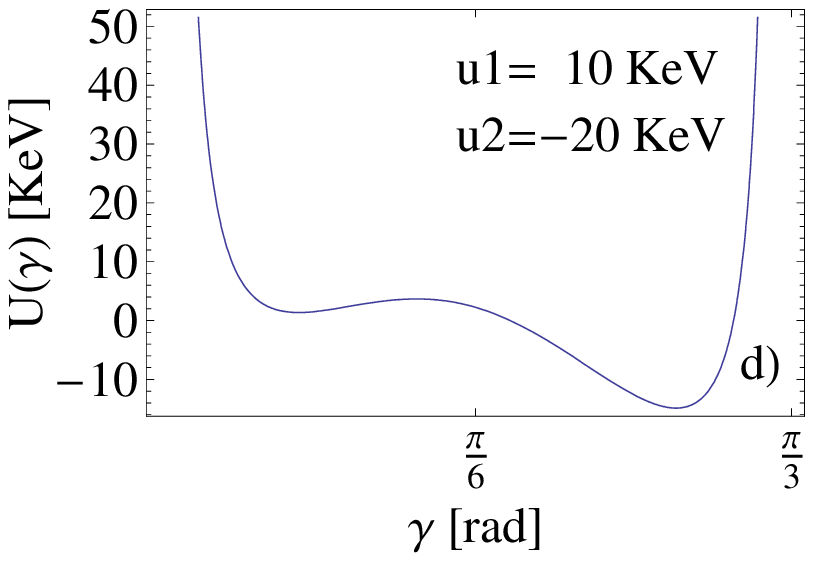}}
\epsfysize=6cm
\centerline{\epsfbox{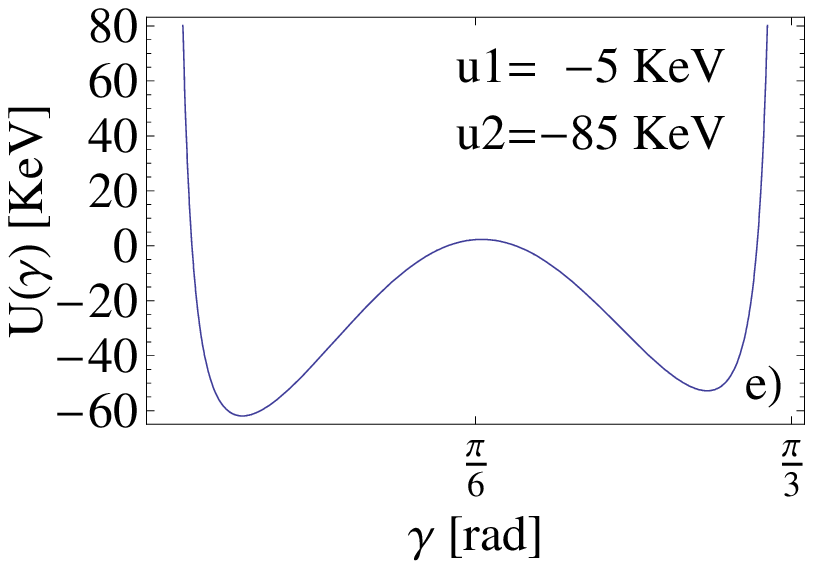}}
\epsfysize=6cm
\centerline{\epsfbox{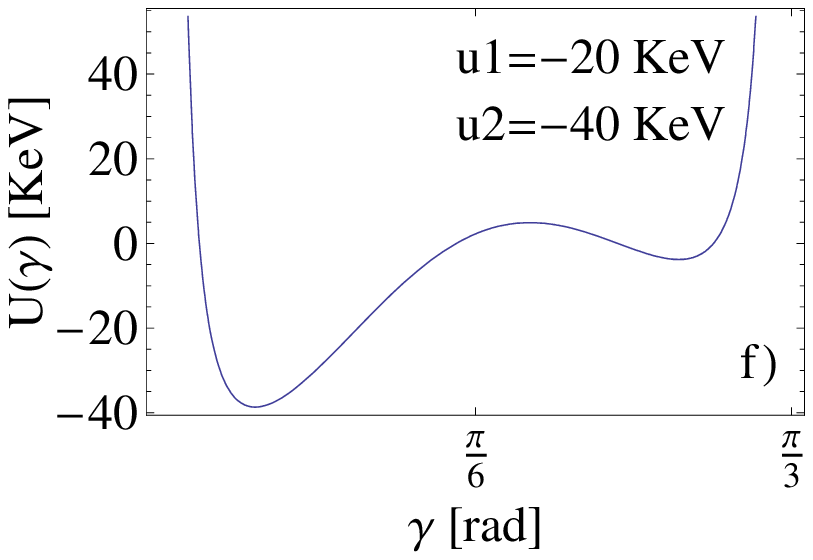}}
\caption{The $\gamma$ potentials (3.8), corresponding to the parameters $(u_1,u_2)=(10,-20);(-5,-85);(-20,-40) keV$ and $\langle \beta^2\rangle =1$,  are represented as functions of $\gamma$ in panels d), e) and f), respectively. }
\end{minipage}
\end{figure}

\begin{figure}[ht!]
\begin{minipage}[t]{8cm}
\epsfysize=6cm
\centerline{\epsfbox{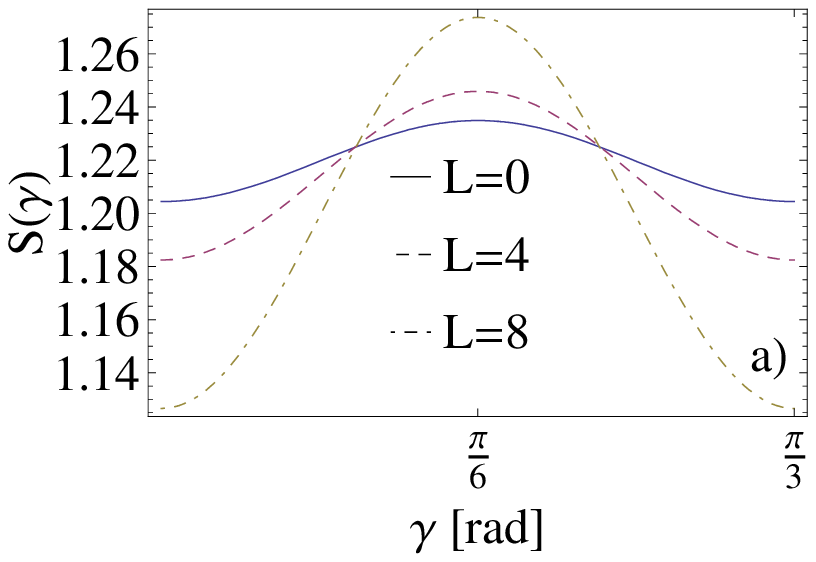}}
\epsfysize=6cm
\centerline{\epsfbox{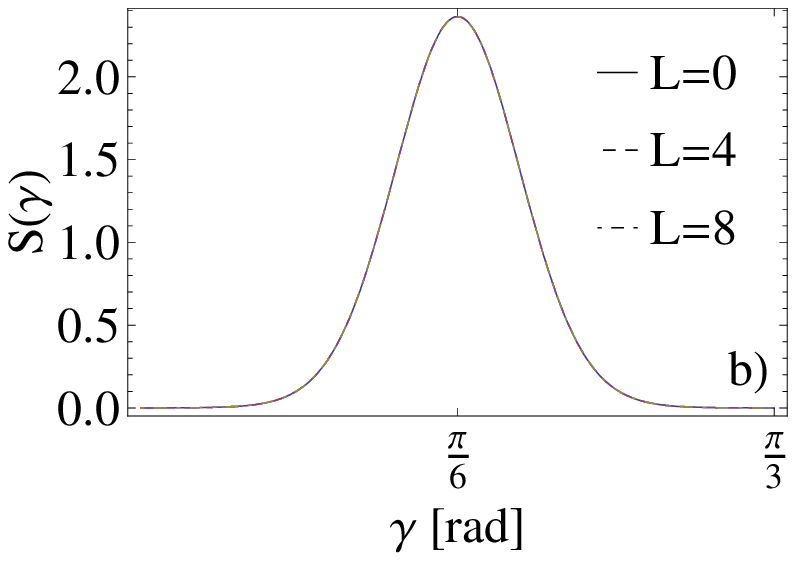}}
\epsfysize=6cm
\centerline{\epsfbox{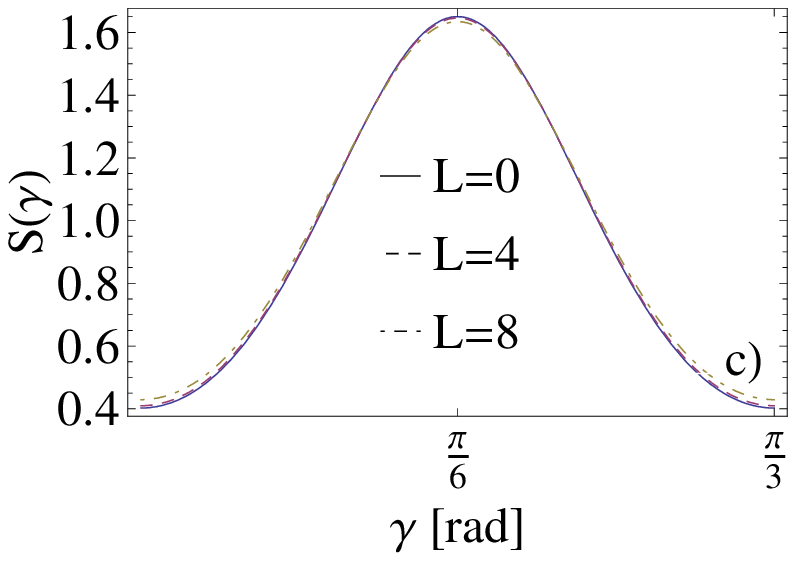}}
\caption{ The  spheroidal functions corresponding to the $\gamma$ potentials from Figs. 6 a), 6 b) and 6 c), are given as functions of $\gamma$ in panels  a), b) and c) respectively.
The functions describe the ground band states of angular momenta 0, 4 and 8, respectively.}
\end{minipage}
\hspace{\fill}
\begin{minipage}[t]{8cm}
\epsfysize=6cm
\centerline{\epsfbox{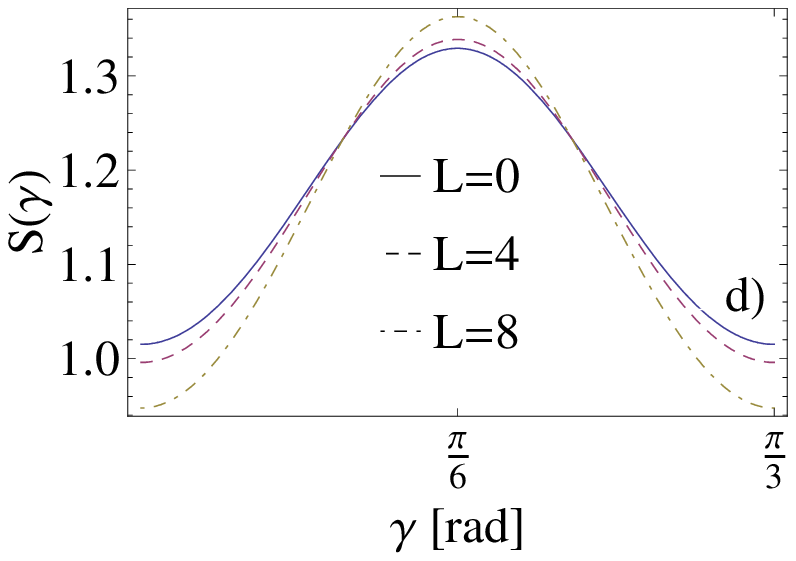}}
\epsfysize=6cm
\centerline{\epsfbox{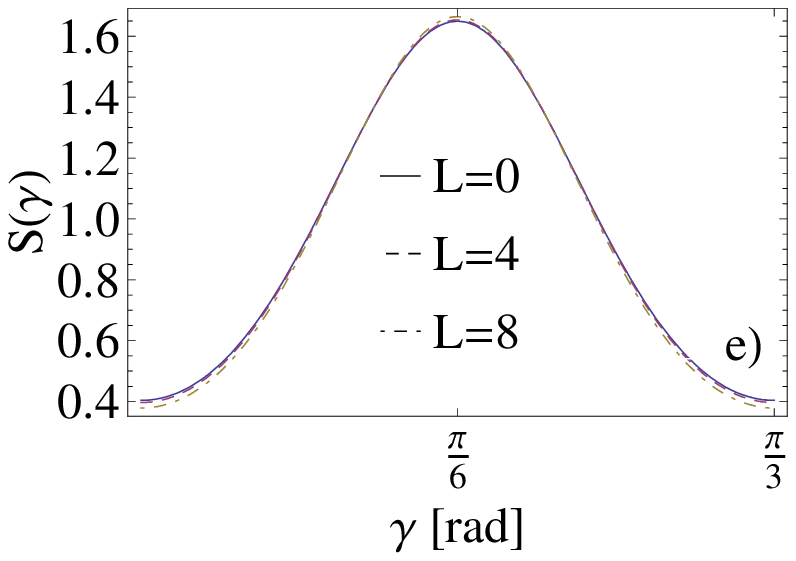}}
\epsfysize=6cm
\centerline{\epsfbox{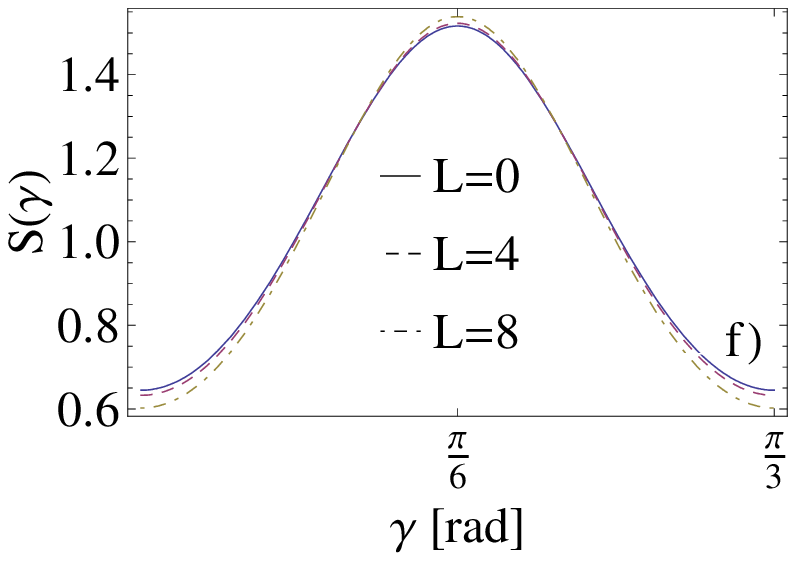}}
\caption{The  spheroidal functions corresponding to the $\gamma$ potentials from Figs. 6 d), 6 e) and 6 f), are plotted versus $\gamma$ in panels d), e) and f) respectively, for three states 
 of angular momenta 0, 4 and  8 belonging to the ground band. }
\end{minipage}
\end{figure}
\clearpage

\begin{figure}[ht!]
\begin{minipage}[t]{8cm}
\epsfysize=5cm
\centerline{\epsfbox{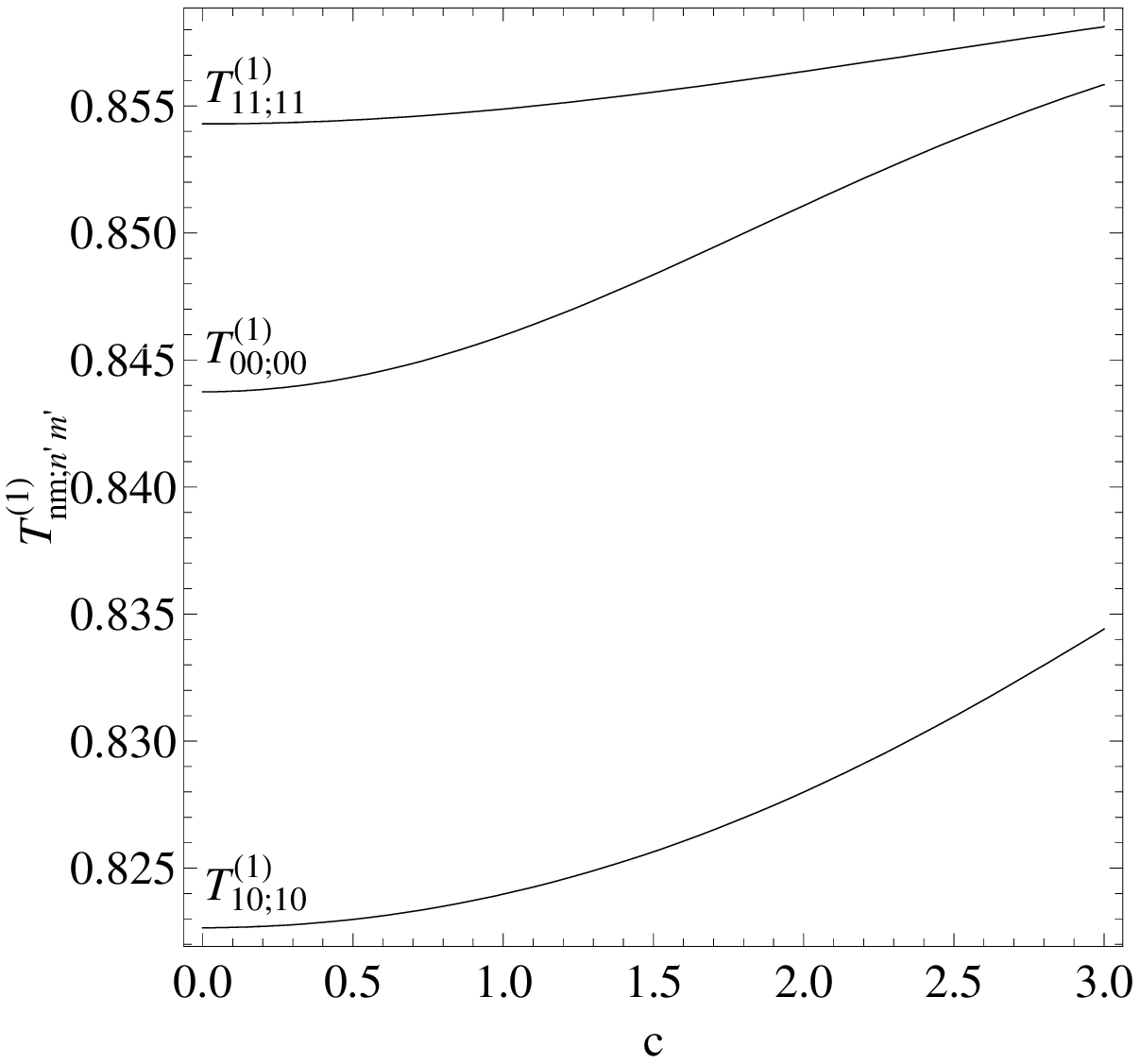}}
\epsfysize=5cm
\centerline{\epsfbox{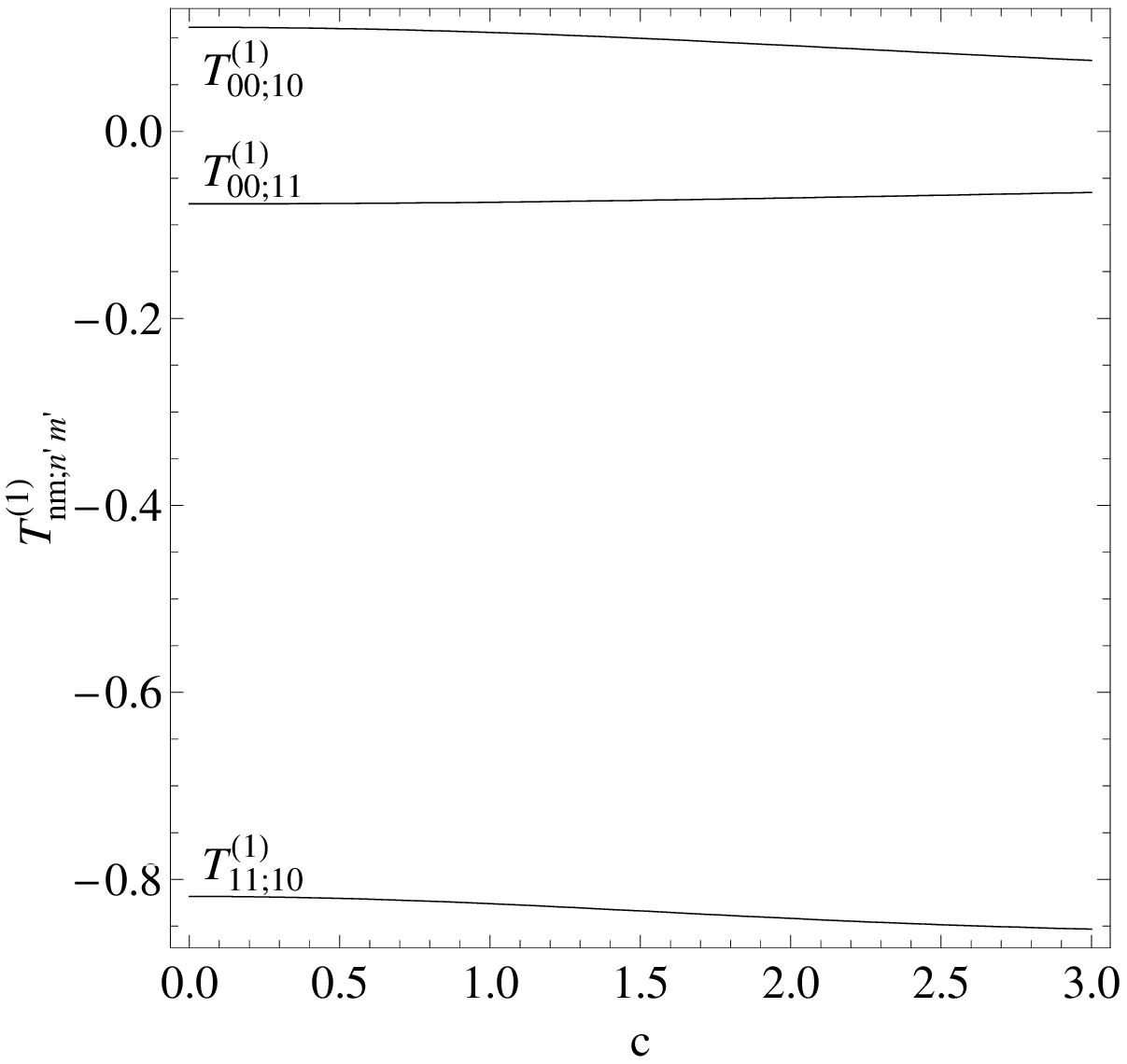}}
\epsfysize=5cm
\centerline{\epsfbox{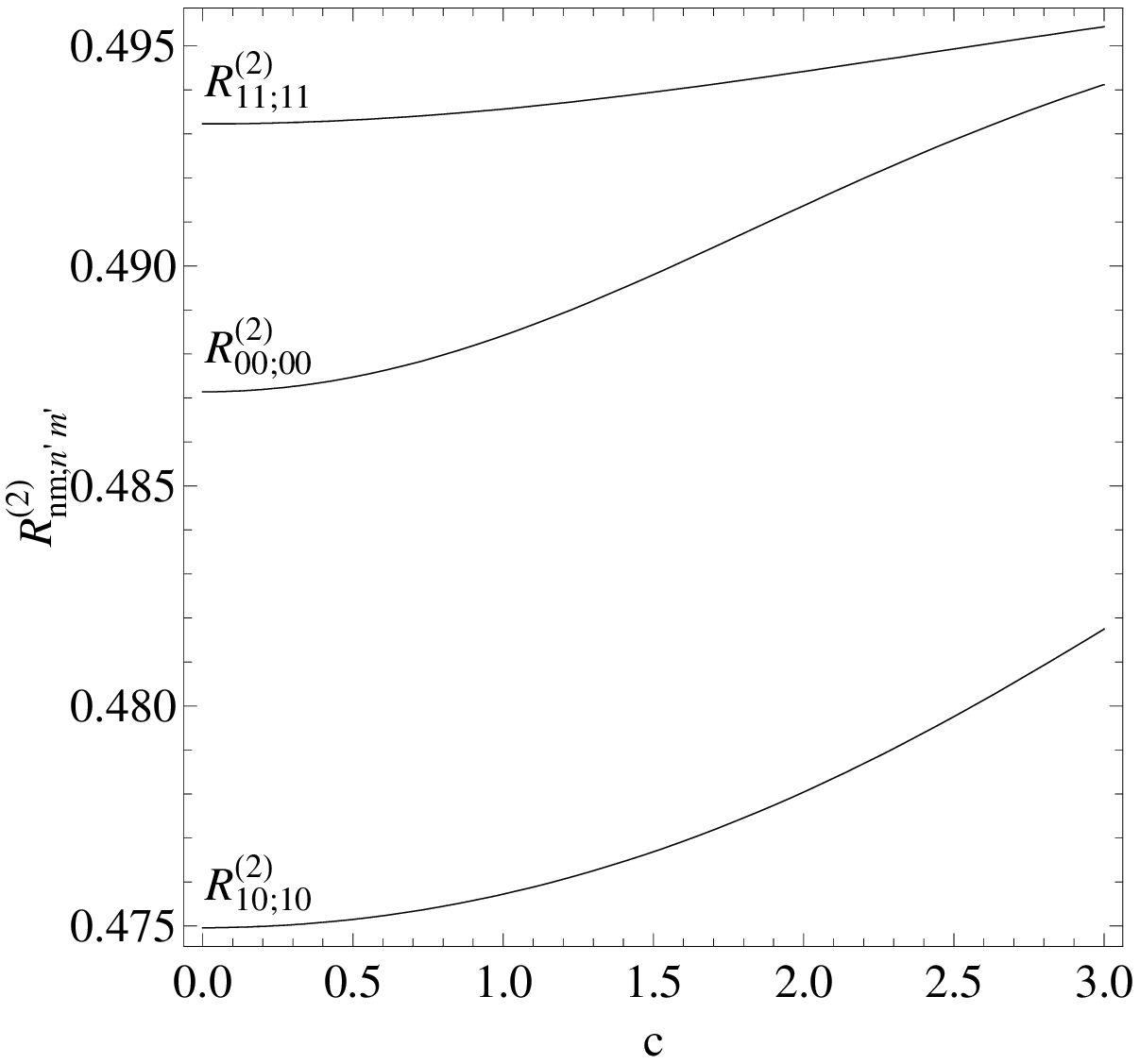}}
\epsfysize=5cm
\centerline{\epsfbox{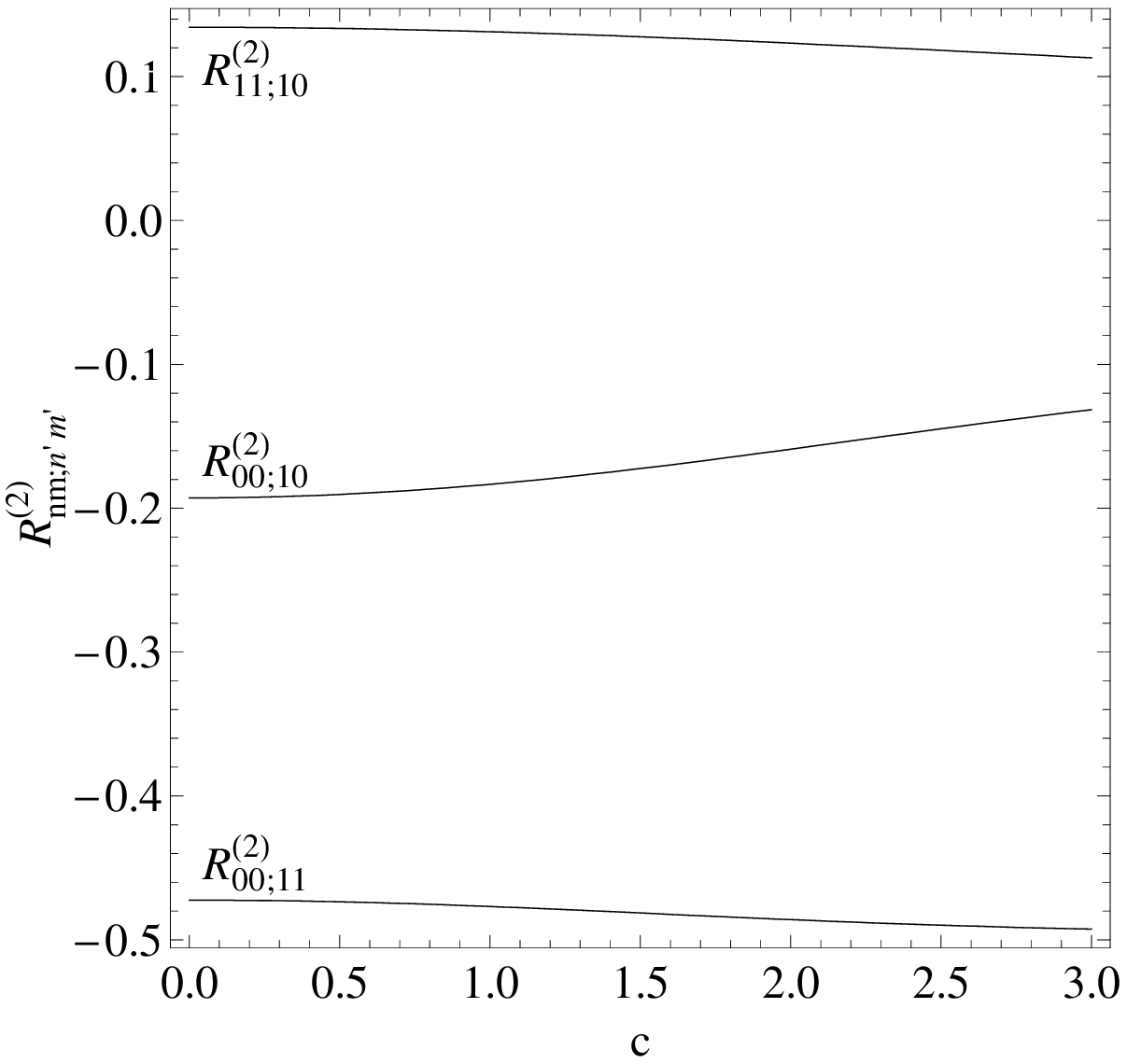}}
\vspace*{-0.3cm}
\caption{ The matrix elements of $\cos\gamma$ ($T^{(1)}_{nm;n'm'}$) and $\sin\gamma$ ($R^{(2)}_{nm;n'm'}$) are plotted as functions of the parameter $c$ involved in Eq.(2.39), for $(nm;n'm')= (11;11), (00;00), (10;10)$ (1st and 3rd panels) $(00;10), (00,11), (11,10)$ (2nd and 4th panels).}
\end{minipage}
\hspace{\fill}
\begin{minipage}[t]{8cm}
\epsfysize=5cm
\centerline{\epsfbox{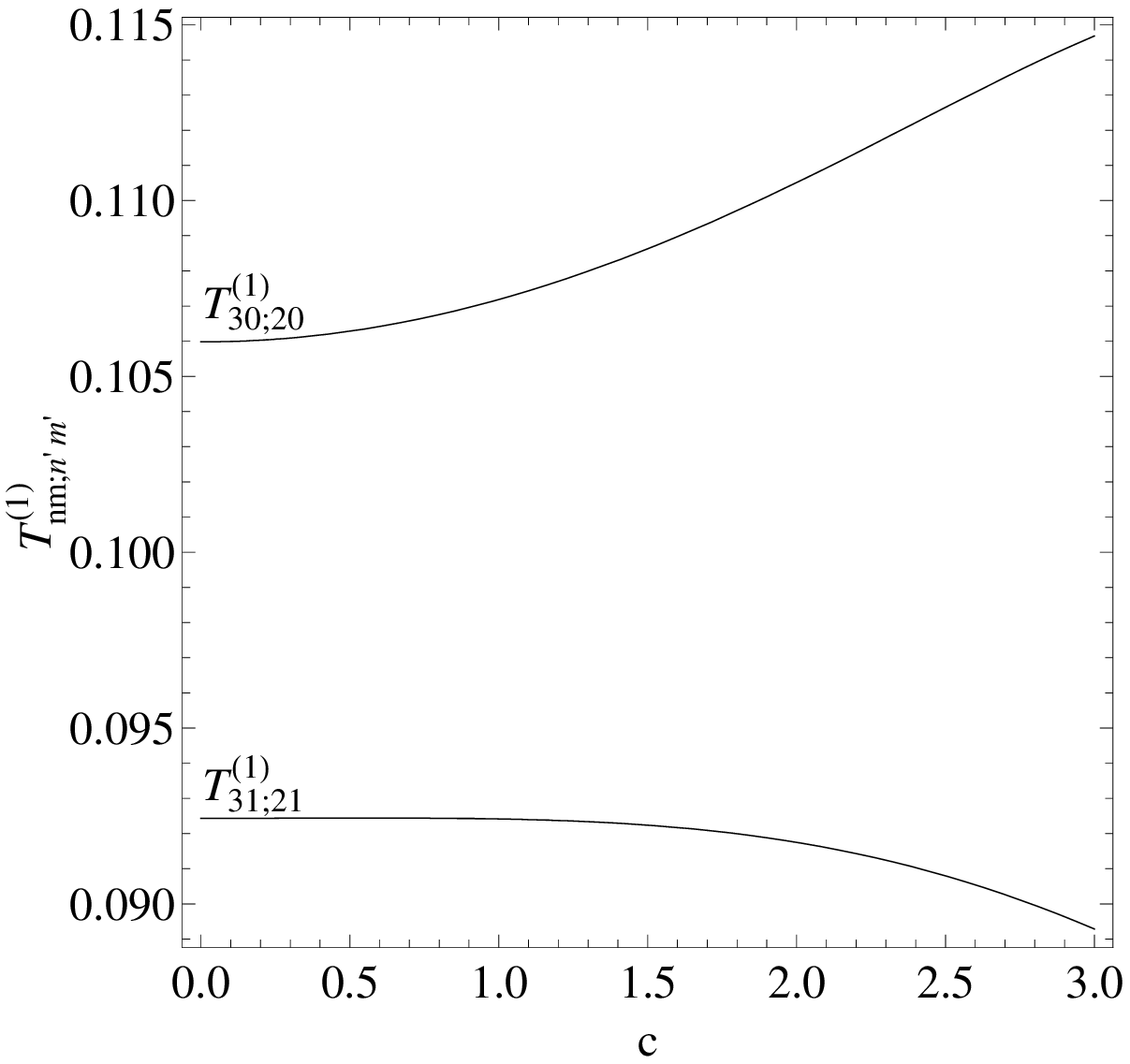}}
\epsfysize=5cm
\centerline{\epsfbox{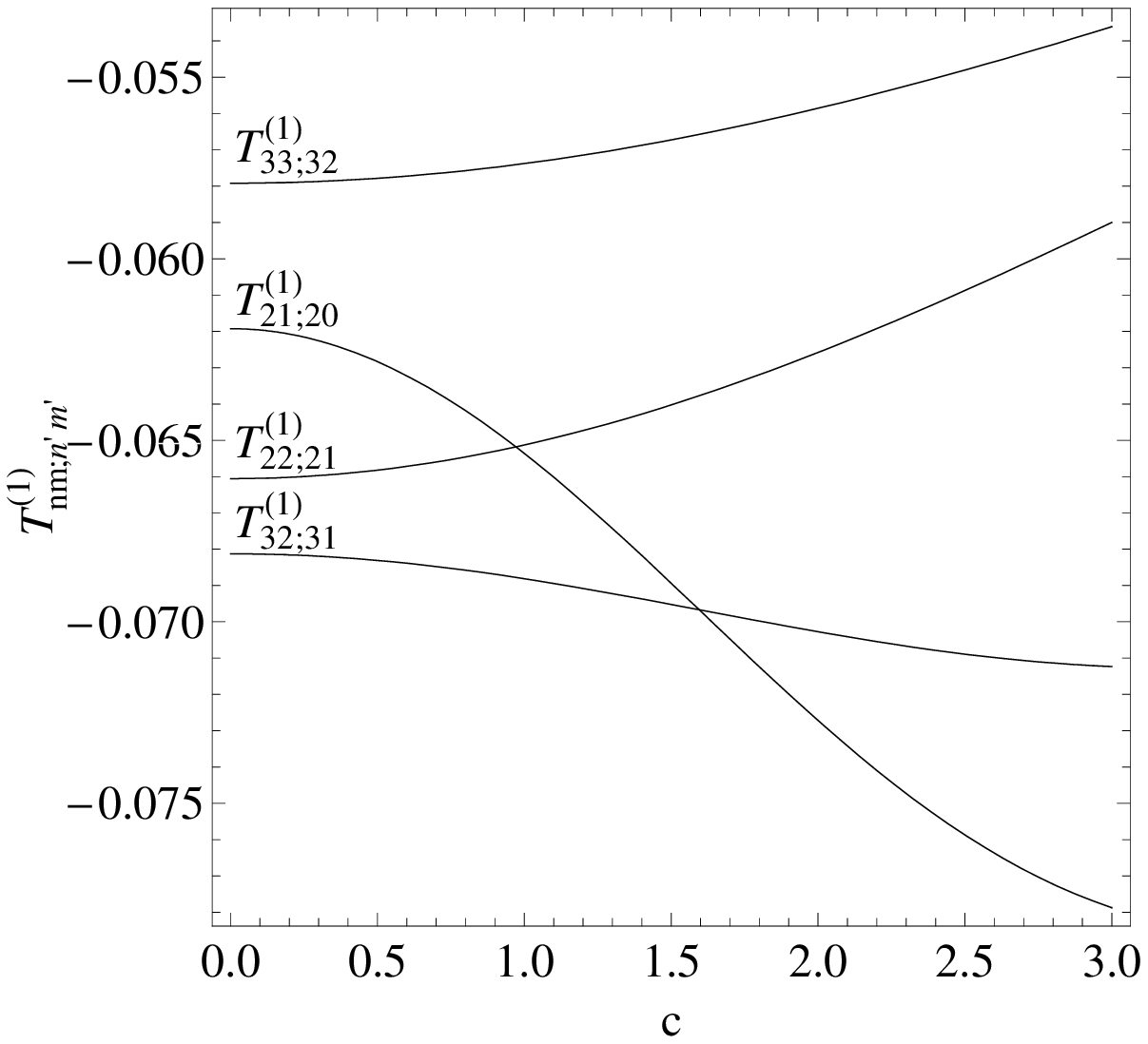}}
\epsfysize=5cm
\centerline{\epsfbox{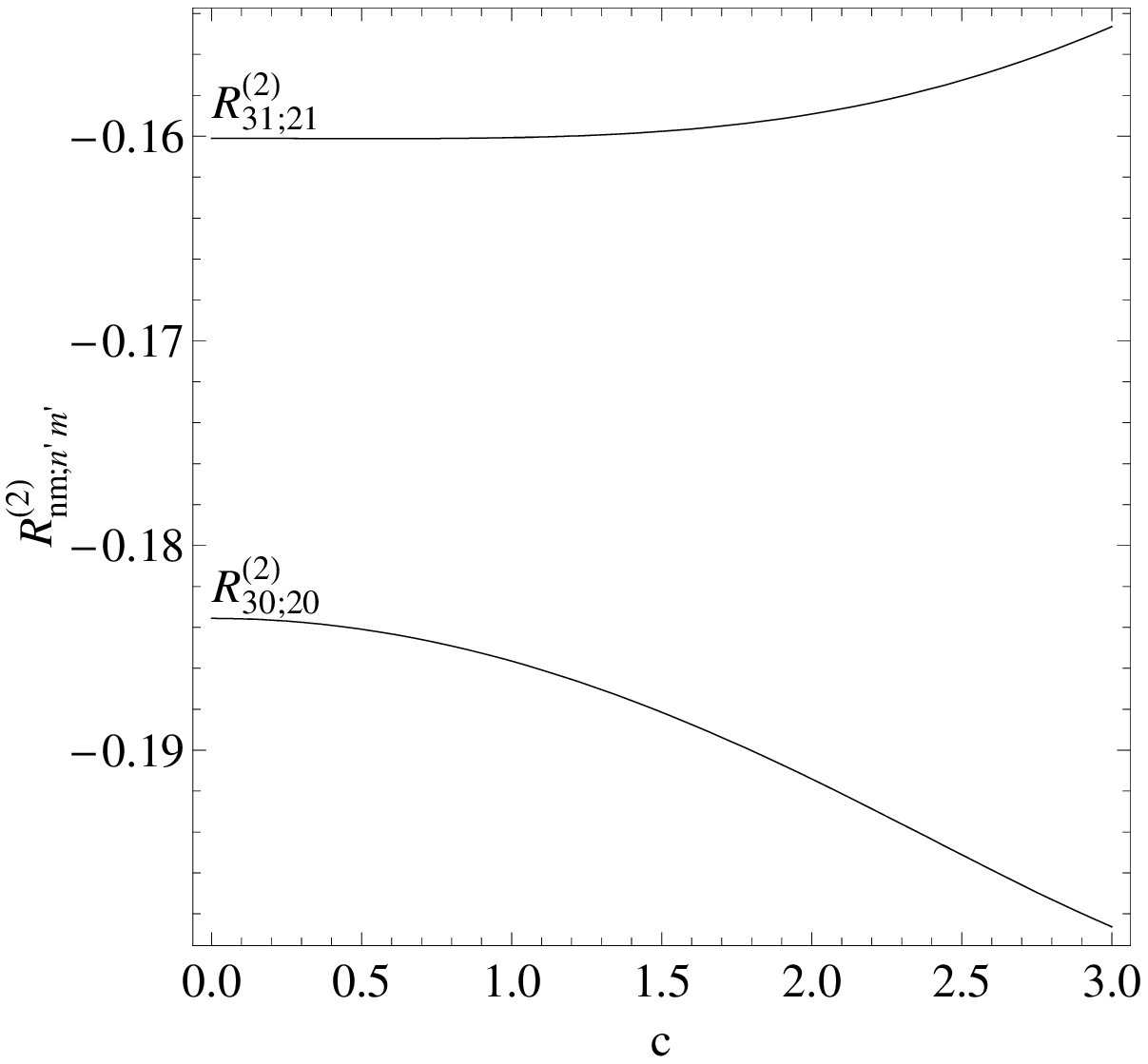}}
\epsfysize=5cm
\centerline{\epsfbox{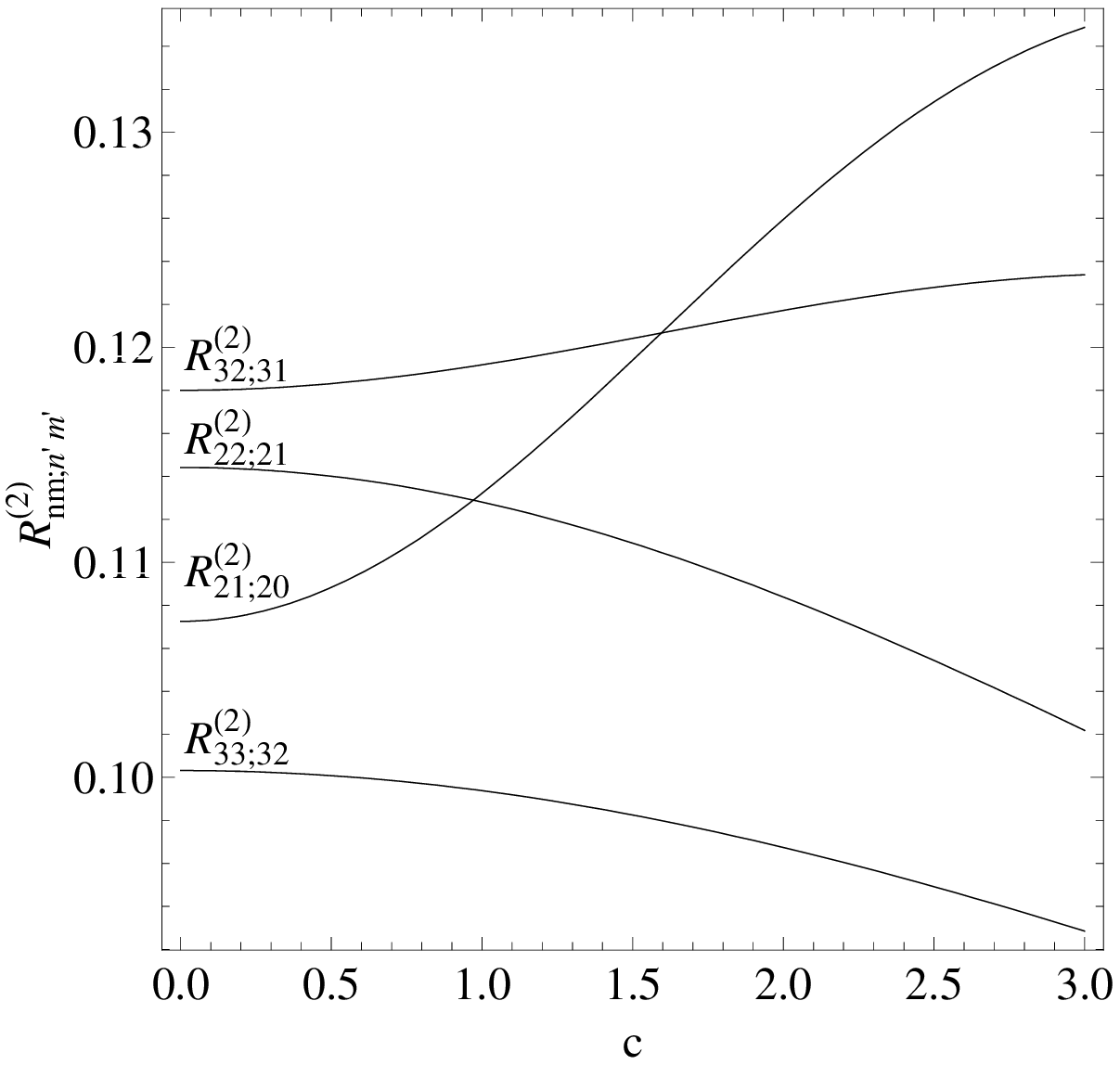}}
\vspace*{-0.3cm}
\caption{The same as in Fig.(10) but for different quantum numbers $(nm;n'm')=
(30;20), (31;21)$ (1st and 3rd panels) $(33,32), (21;20), (22;21), (32; 31)$ (2nd and 4th panels). }
\end{minipage}
\end{figure}
\clearpage

\pagebreak

\pagebreak

\section{Numerical results}
\label{sec:level6}
The formalism described in the previous Section has been applied for $^{150}$Nd, $^{154}$Gd and $^{192}$Os. 
The choice is justified by the values of the ratio of the excitation energies for the first two excited states in the ground band. Indeed, these are 2.93, 3.015 and 2.82 respectively and therefore they are expected to have the features of X(5) symmetry. As we stated in Introduction the present formalism is close to the X(5) symmetry. Indeed, it goes to $X(5)$ in the limit $\gamma \to0$ and, on the other hand, the spheroidal function equation has been derived by expanding the gamma depending terms in the initial Hamiltonian in terms of $\sin3\gamma$. However, it brings two new things namely  i) the wave function  is periodic in $\gamma$, the matrix elements of $\gamma$ depending functions being performed with the measure $|\sin3\gamma|d\gamma$ and ii)  the factor function describing the $\beta$ degree of freedom satisfies a Schr\"{o}dinger equation involving the Davidson's potential. Numerical calculations will show us what are the quantitative corrections to the $X(5)$ picture
brought by curing the drawbacks of the preceding approaches.

The $\gamma$ potentials for the three nuclei have been chosen from those given in Fig.6. Indeed, the potentials $^{154}$Gd, $^{150}$Nd and $^{192}$Os are those from Figs. 6 a), c) and b) respectively. The spheroidal functions corresponding to these potentials are represented in Figs 8 a), 8 c) and 8 b), respectively. We varied the shapes of gamma potential and made the option for that one which yields a good agreement between the  calculated B(E2) values associated to the transitions from gamma to ground band and the corresponding experimental data.

The energy levels in the three bands are given in units of $E_{2^+_g}$ and therefore we need only the ratio of the parameters  $A$ and $F$ involved in the energy expression given by Eq. (3.10). The term $E_0$ is not depending on the quantum number defining the states but it is considered to depend on band. In our calculation $E_0$ for  ground and beta bands  are taken to be equal and therefore they do not affect the relative energies in the two bands, while for gamma band was fixed  so that the head state energy  is recovered.
The parameters defining the transition operator have been fixed by fitting the B(E2) values for the transitions $2^+_g\to 0^+_g$ and $2^+_{\gamma}\to 0^+_{g}$. The results obtained in this manner are collected in Table I.

\begin{table}
\begin{tabular}{|c|c|c|c|}
\hline
                  &  $\hskip0.5cm$$^{150}$Nd $\hskip0.5cm$  & $\hskip0.5cm$ $^{154}$Gd $\hskip0.5cm$  &  $\hskip0.2cm$ $^{192}$Os $\hskip0.5cm$  \\
\hline
$u_1[keV]$        &   100         &   -1          &      5           \\
$u_2[keV]$        &  1100         &   -1          &     85            \\
\hline
$E_0 [keV]$       &-27.696        &  -222.243     &      9.635        \\
                  &-3.545         &  -12.892      &      0.367         \\
\hline
$F/A$             &0.532          & 17.728        &   0.191$\cdot10^{-3}$\\
                  &0.0328         &1.027          & 2.896$\cdot 10^{-3}$   \\
\hline
$t [efm^2]$       &147.889        &124.065        &90.583\\
                  &161.909        &135.831        &99.174 \\
\hline   
$t_1 [efm^2]$     & 209.443       &184.933        &120.976 \\
                  &228.706        &219.473        &138.962  \\
\hline
$t_2 [efm^2]$     &100.178        & 106.240       &50.077\\
                  &116.624        &156.629        &71.225 \\
\hline
\end{tabular}
\caption{The parameters $u_1, u_2, E_0, F/A$, involved in the energy expression (3.10), calculated  by the method described in the text, are given for  $^{150}$Nd,  $^{154}$Gd, $^{192}$Os. Also we give the values for the parameters $t$ and  $t_1, t_2$ involved in the  harmonic and anharmonic transition operators, respectively. They were obtained by fitting the B(E2) values for the transition $2^+_g\to 0^+_g$ if the harmonic transition operator is used, and
for the transitions $2^+_g\to 0^+_g$ and $2^+_{\gamma}\to 0^+_g$ for the anharmonic transition operator. } 
\end{table}

Now let us proceed at describing separately the results for each of the three nuclei considered  here.

\begin{table}
\vspace{0.5cm}
\begin{tabular}{|c|cccc|}
\hline
State&Exp.&X(5)&ISW&D\\
\hline
$2_{g}^{+}$&1&1&1&1\\
$4_{g}^{+}$&2.93&2.90&2.93&2.93\\
$6_{g}^{+}$&5.53&5.43&5.53&5.53\\
$8_{g}^{+}$&8.68&8.48&8.70&8.73\\
$10_{g}^{+}$&12.28&12.03&12.42&12.50\\
$12_{g}^{+}$&16.27&16.04&16.66&16.82\\
\hline
$0_{\beta}^{+}$&5.19&5.65&5.30&4.20\\
$2_{\beta}^{+}$&6.53&7.45&7.05&5.45\\
$4_{\beta}^{+}$&8.74&10.69&10.24&7.65\\
$6_{\beta}^{+}$&11.84&14.75&14.27&10.64\\
\hline
$2_{\gamma}^{+}$&8.16&8.16&8.16&8.16\\
$3_{\gamma}^{+}$&9.22&8.93&9.03&9.35\\
$4_{\gamma}^{+}$&10.39&9.83&10.08&10.70\\
\hline
\end{tabular}
\caption{Excitation energies of some states from ground, beta and gamma  bands 
of $^{150}$Nd, given in units of the $2^+_g$ energy, obtained with three different approaches, $X(5)$, 
infinite square well potential ($ISW$) for $\beta$ and spheroidal functions formalism for $\gamma$, 
Davidson's $\beta$ potential plus spheroidal
functions method for $\gamma$ ($D$), are compared with the corresponding experimental data.}
\end{table}

The calculated energies for $^{150}$Nd are listed in Table II. The results of present paper are given in the columns $D$ and $ISW$. They have been obtained by using a spheroidal function description in the $\gamma$ variable while for  the $\beta$ potential, an infinite square well ($ISW$) and the Davidson's potentials (D), have been alternatively used. These results are compared with the corresponding experimental data \cite{Kruc,Mateo}, listed in the column headed by $Exp.$, as well as with the theoretical results obtained within the $X(5)$ approach. If we enlarge the list for each band the deviations of predictions from the corresponding experimental data are increasing  functions of angular momentum.

\pagebreak
\begin{table}
\vspace{0.5cm}
\begin{tabular}{|ccccccc|}
\hline
$B(E2;J^+_k\to J'^+_{k'}$&Exp.&X(5)&\multicolumn{2}{c}{ISW}&\multicolumn{2}{c|}{D}\\
\hline
$2_{g}^{+}\rightarrow 0_{g}^{+}$&115&115&115&115&115&115\\
$4_{g}^{+}\rightarrow 2_{g}^{+}$&182&184&184&177&197&182\\
$6_{g}^{+}\rightarrow 4_{g}^{+}$&210&228&228&210&266&226\\
$8_{g}^{+}\rightarrow 6_{g}^{+}$&278&262&262&233&334&260\\
$10_{g}^{+}\rightarrow 8_{g}^{+}$&204&288&289&249&394&387\\
\hline
$2_{\beta}^{+}\rightarrow 0_{\beta}^{+}$&114&92&92&91&148&121\\
$4_{\beta}^{+}\rightarrow 2_{\beta}^{+}$&170&138&138&136&210&167\\
\hline
$0_{\beta}^{+}\rightarrow 2_{g}^{+}$&39&72&72&43&105&58\\
$2_{\beta}^{+}\rightarrow 0_{g}^{+}$&1.2&2.4&2.4&0.4&0.6&0.1\\
$2_{\beta}^{+}\rightarrow 2_{g}^{+}$&9&10&10&4&9&3\\
$2_{\beta}^{+}\rightarrow 4_{g}^{+}$&17&42&42&26&69&36\\
$4_{\beta}^{+}\rightarrow 2_{g}^{+}$&0.12&0.56&1&0.01&0.61&4.6\\
$4_{\beta}^{+}\rightarrow 4_{g}^{+}$&7&7&7&3&4&0.3\\
$4_{\beta}^{+}\rightarrow 6_{g}^{+}$&70&32&32&18&51&22\\
\hline
$2_{\gamma}^{+}\rightarrow 0_{g}^{+}$&3.0&3.4&0.6&3&0.6&3\\
$2_{\gamma}^{+}\rightarrow 2_{g}^{+}$&5.4&5.1&0.9&4.7&1&5.0\\
$2_{\gamma}^{+}\rightarrow 4_{g}^{+}$&2.6&0.3&0.05&0.3&0.05&0.3\\
$4_{\gamma}^{+}\rightarrow 2_{g}^{+}$&0.9&2.3&0.4&2.1&0.4&2.3\\
$4_{\gamma}^{+}\rightarrow 4_{g}^{+}$&3.9&7.3&1.3&6.8&1.5&8.2\\
\hline
\end{tabular}
\caption{The B(E2) values for $^{150}$Nd  calculated in three different formalisms, X(5), ISW, B, and the corresponding experimental data are given in units of $e^2fm^4\times10^2$. The results of the present work are obtained by using spheroidal functions for $\gamma$ and alternatively an infinite square well potential ($ISW$) and  Davidson's potential ($D$) for the variable $\beta$. The results for $X(5)$ formalism are taken from Ref.\cite{Kruc}. In the first columns headed by $ISW$ and $D$ respectively, are the results obtained with a harmonic quadrupole transition operator. In the second columns $ISW$ and $D$ we give the results obtained with an anharmonic quadrupole transition operator. Calculations made with the $X(5)$ formalism correspond to a harmonic transition operator.}
\end{table}

The intraband ground band and beta band transitions as well the gamma to ground and beta to ground transitions were calculated by using both a harmonic and an anharmonic structure for the transition operators with the strength parameters from Table I. The final results are those from Table III. In the mentioned Table we give also the corresponding experimental data,  taken from Ref.\cite{Kruc,Mateo}, as well as the results provided by the X(5) formalism. It is interesting to remark that except for the gamma to ground transitions, the harmonic approach of the $ISW$ calculations provides identical results with the X(5) calculations.
We may say that the agreement with the data is improved by adding the anharmonic effects for some transitions  but for some other the discrepancies are increased. In an overall analysis, the agreement is improved by anharmonicities.

\begin{table}
\vspace{0.5cm}
\begin{tabular}{|r|cccccc|}
\hline
State&Exp.&X(5)&ISW&D&CSM& CSM2\\
\hline
$2_{g}^{+}$&1&1&1&1&1&1\\
$4_{g}^{+}$&3.015&2.90&3.015&3.015&3.110&2.932\\
$6_{g}^{+}$&5.83&5.43&5.84&5.83&6.105&5.612\\
$8_{g}^{+}$&9.30&8.48&9.39&9.35&9.835&8.936\\
$10_{g}^{+}$&13.30&12.03&13.55&13.81&14.188&12.850\\
\hline
$0_{\beta}^{+}$&5.53&5.65&4.12&3.35&5.662&5.335\\
$2_{\beta}^{+}$&6.63&7.45&5.71&4.51&6.413&6.025\\
$4_{\beta}^{+}$&8.51&10.69&8.70&6.68&8.101&7.714\\
$6_{\beta}^{+}$&11.10&14.75&12.63&9.75& 10.626&10.325\\
$8_{\beta}^{+}$&14.27&19.44&17.36&13.58&13.879&13.756\\
$10_{\beta}^{+}$&17.83&24.69&22.79&18.15&17.782&17.916\\
\hline
$2_{\gamma}^{+}$&8.10&8.10&8.10&8.10&7.771&7.817\\
$3_{\gamma}^{+}$&9.16&8.86&9.00&8.89&9.047&8.528\\
$4_{\gamma}^{+}$&10.27&9.75&9.87&10.33&10.073&9.514\\
$5_{\gamma}^{+}$&11.64&10.75&11.01&11.74&11.301&10.604\\
$6_{\gamma}^{+}$&13.05&11.86&12.31&13.31&12.778&12.002\\
$7_{\gamma}^{+}$&14.71&13.07&13.75&15.02&14.386&13.421\\
\hline
\end{tabular}
\caption{The same as in Table II but for $^{154}$Gd.}
\end{table}
\clearpage

Although $^{154}$Gd is a deformed nucleus, having the quadrupole deformation $\approx 0.25$ \cite{Lala}, the authors of Ref.\cite{Clar1} consider it as a good candidate for the critical point in a phase transition which takes place along the chain of the Gd isotopes. This shape transition has been studied recently by two of us (A.A.R. and A. F.) in Ref.\cite{Rad055} within the Coherent State Model (CSM). Here the results obtained through $ISW$ and $D$ formalism are compared to those obtained within other phenomenological approaches like X(5), CSM, and CSM2. The CSM2 differs from CSM by the model states for the beta band.
A full list of references concerning CSM may be found in Ref.\cite{Rad5}.
The results for X(5) are taken from Ref.\cite{Bij} with a suitable scaling when one passes from $^{152}$Sm to $^{154}$Gd.
We notice that the $D$ formalism provides energies which are closest to the experimental data \cite{Lipa,Giri,Rei}.
Concerning the B(E2) values, from Table V one notices that $D$ and $ISW$ formalisms are reproducing better the data \cite{Lipa,Giri} for the intraband ground and beta bands transitions. As regards the beta to ground transitions the X(5) formalism yields an overall better agreement with the data. Also, we note that the approach $D$ is describing quite well the transitions $J^+_{\beta}\to (J+2)^+_g$ but its predictions for $J^+_{\beta}\to J^+_g$ are smaller than the corresponding data by a factor of about 10. By contrary the theoretical transitions $J^+_{\beta}\to (J-2)^+_g$ are larger by a factor 10 to 50 than the experimental results. The best description for the interband transitions collected in Table V is provided by the $ISW$ approach. 
The anharmonic term of the transition operator brings important contribution to both the intra and interband transitions.

\begin{table}
\vspace{0.5cm}
\begin{tabular}{|ccccccccccc|}
\hline
Transition&Exp.&X(5)&\multicolumn{2}{c}{ISW}&\multicolumn{2}{c}{D}&\multicolumn{2}{c}{CSM}&\multicolumn{2}{c|}{CSM2}\\
\hline
$2_{g}^{+}\rightarrow 0_{g}^{+}$&77.3&77.3&77.3&77.3&77.3&77.3&77.3&77.3&77.3&77.3\\
$4_{g}^{+}\rightarrow 2_{g}^{+}$&117.8&123.5&123.7&117.8&132.4&117.8&114.4&114.9&117.2&117.3\\
$6_{g}^{+}\rightarrow 4_{g}^{+}$&138.2&153&153.4&138.9&179.1&138.9&133.1&133.9&139.7&139.9\\
$8_{g}^{+}\rightarrow 6_{g}^{+}$&152.6&176.1&176.3&152.8&225&152.2&148.7&148.4&159.4&159.8\\
$10_{g}^{+}\rightarrow 8_{g}^{+}$&173.1&193.8&194.6&162.4&265.6&160&163.7&163.8&178.4&178.9\\
\hline
$2_{\beta}^{+}\rightarrow 0_{\beta}^{+}$&49&61.7&61.5&61.3&99.2&72.7&103.8&67.1&102.8&103.8\\
$4_{\beta}^{+}\rightarrow 2_{\beta}^{+}$&122&92.9&92.8&91.2&141.1&99.3&152.4&99.5&150.9&152.4\\
$6_{\beta}^{+}\rightarrow 4_{\beta}^{+}$&111&113.6&113.6&109.5&152.4&107.5&175.6&116.2&174&175.6\\
\hline
$0_{\beta}^{+}\rightarrow 2_{g}^{+}$&25.8&48.8&48.2&26&70.6&26.5&1(-3)&1.32&19.9&20.0\\
$2_{\beta}^{+}\rightarrow 2_{g}^{+}$&4.0&7.0&6.4&2.1&6.1&0.44& 2(-4)&0.34&4.13&4.16\\
$2_{\beta}^{+}\rightarrow 4_{g}^{+}$&11.9&27.9&28.3&15.4&46.4&15.9& 3(-3)&0.76&12.98&13.08\\
$4_{\beta}^{+}\rightarrow 2_{g}^{+}$&0.35&0.54&0.73&0.06&0.4&5.4&  0  &0.21&2.17 &2.19\\
$4_{\beta}^{+}\rightarrow 4_{g}^{+}$&3.8&4.83&4.72&1.29&2.8&0.044& 3(-3)&0.33&3.44&3.47\\
$4_{\beta}^{+}\rightarrow 6_{g}^{+}$&12&21.5&21.6&10.7&34.3&8.16& 8(-3)&0.68&13.42&13.53\\
$6_{\beta}^{+}\rightarrow 4_{g}^{+}$&0.27&0.50&0.50&0.18&7&17.3&  1(-4)&0.17&1.70&1.71\\
\hline
\end{tabular}
\caption{The same as in Table III but for $^{154}$Gd. The notation $k(-m)$ stands for the number $k\cdot 10^{-m}$.}
\end{table}
\clearpage

\begin{table}
\vspace{0.5cm}
\begin{tabular}{|r|cccccccccl|}
\hline
$L_{i},L_{f};L_{i}^{'},L{f}^{'}$&Exp.&X(5)&\multicolumn{2}{c}{ISW}&\multicolumn{2}{c}{D}&\multicolumn{2}{c}{CSM}&\multicolumn{2}{c|}{CSM2}\\
\hline
$4_{g},2_{g};2_{g},0_{g}$&1.52&1.60&1.60&1.52&1.71   &1.52&1.48   &1.48&1.52   &1.52\\
$6_{g},4_{g};4_{g},2_{g}$&1.17&1.24&1.24&1.18& 1.35   &1.18&1.16   &1.16& 1.19   &1.19\\
$8_{g},6_{g};6_{g},4_{g}$&1.10&1.15&1.15&1.10& 1.26   &1.10&  1.12  &1.11&1.14    &1.14\\
$10_{g},8_{g};8_{g},6_{g}$&1.13&1.10&1.10&1.06&1.18    &1.06& 1.10   &1.10&1.119    &1.113\\
\hline
$2_{\gamma},0_{g};2_{\gamma},2_{g}$&0.468&0.666&0.654    &0.642&0.618     &0.594&0.509&0.468&0.501   &0.468\\
$2_{\gamma},4_{g};2_{\gamma},2_{g}$&0.144&0.052&0.053    &0.054&0.054     &0.057&0.087&0.131&0.068    &0.069\\
$3_{\gamma},2_{g};3_{\gamma},4_{g}$&1.006&2.368&2.335    &2.274&2.208      &2.096&1.302&0.975&1.432    &1.289\\
$4_{\gamma},2_{g};4_{\gamma},4_{g}$&0.148&0.315& 0.311   &0.304& 0.286     &0.273&0.159&0.126&0.152    &0.123\\
$4_{\gamma},6_{g};4_{\gamma},4_{g}$&0.27&0.088& 0.090    &0.092& 0.092     &0.097& 0.377&0.377&0.115      &0.117\\
$5_{\gamma},4_{g};5_{\gamma},6_{g}$&0.744&1.667& 1.655   &1.619& 1.560     &1.486&0.657&0.401&0.786     &0.665\\
$6_{\gamma},4_{g};6_{\gamma},6_{g}$&0.081&2.333& 0.25    &0.246& 0.229     &0.219& 0.081&0.046&0.076     &0.050\\
$2_{\gamma},2_{\beta};2_{\gamma},2_{g}$&1.00&0.032& 0.054&0.108& 0.061     &0.145&1.206&0.751&2.238   &0.048\\
\hline
$2_{\beta},0_{g};2_{\beta},2_{g}$&0.123&0.429& 0.257     &0.067&0.069      &1.716&0.0&0.561&0.538    &0.475\\
$2_{\beta},4_{g};2_{\beta},2_{g}$&2.76&4.00&  4.45       &7.20& 7.6      &36& 13.24&2.257& 3.141    &3.141\\
$4_{\beta},2_{g};4_{\beta},4_{g}$&0.086&0.112& 0.155     &0.046&0.145      &121&0.001&0.625& 0.630  &0.630\\
$4_{\beta},6_{g};4_{\beta},4_{g}$&2.63&4.45&  4.57       &8.30& 12      &184&2.484&2.073& 3.896     &3.896\\
$6_{\beta},4_{g};6_{\beta},6_{g}$&0.08&&      0.13       &0.22&   2475     &292& 0.0&0.071& 0.555   &0.555\\
\hline
$2_{\beta},0_{g};2_{\beta},0_{\beta}$&0.008&0.049&0.027   &0.002&  0.004    &0.568&0.0&0.0028&0.022    &0.019\\
$4_{\beta},2_{g};4_{\beta},2_{\beta}$&0.0025&0.0058&0.0079&0.0006& 0.0029    &0.530&0.0&0.0021& 0.014   &0.014\\
$6_{\beta},4_{g};6_{\beta},4_{\beta}$&0.0024&0.0044&0.0044&0.0017& 0.045    &0.676& 0.0&0.0015& 0.010   &0.010\\
$8_{\beta},6_{g};8_{\beta},6_{\beta}$&0.006&&       0.003 &0.002&  0.204    &2.147& 0.977&0.0011& 0.007    &0.007\\
\hline
\end{tabular}
\caption{Calculated branching ratios $B(E2;L_{i}^{+}\rightarrow L_{f}^{+})/B(E2;L_{i}^{'+}\rightarrow L_{f}^{'+})$ (denoted by 
$L_{i},L_{f};L_{i}^{'},L{f}^{'}$ ) for some interband as well intraband transitions in $^{154}$Gd are compared with the
corresponding experimental data. The results of this paper, labeled by ISW and D, were obtained by using spheroidal functions for $\gamma$ and for the $\beta$ variable an infinite square well and the Davidson's potentials, respectively.
Results from the first columns headed by $ISW$ and $D$ were obtained by using a harmonic transition quadrupole operator while in the second columns are listed the results corresponding to an anharmonic transition operator.}  
\end{table}
\clearpage
Some branching ratios of the intraground band as well as of the interband transitions are given in Table VI. The failure of the $D$ approach to describe these branchings for the $\beta\to g$ transitions is noticeable. The reason consists in the fact that within this formalism the predictions for the transitions $J^+_{\beta}\to J^+_g$ are too small comparing them with the experimental data. The other ratios are described reasonable well by all theoretical models.

\begin{table}
\vspace{0.5cm}
\begin{tabular}{|ccccc|}
\hline
Starea&Exp.&X(5)&ISW&D\\
\hline
$2_{g}^{+}$&1&1&1&1\\
$4_{g}^{+}$&2.82&2.90&2.90&2.90\\
$6_{g}^{+}$&5.29&5.43&5.43&5.43\\
$8_{g}^{+}$&8.30&8.48&8.48&8.48\\
$10_{g}^{+}$&11.75&12.03&12.03&12.01\\
$12_{g}^{+}$&15.60&16.04&16.04&16.01\\
\hline
$0_{\beta}^{+}$&4.65&5.65&5.65&4.44\\
$2_{\beta}^{+}$&5.48&7.45&7.45&5.73\\
\hline
$2_{\gamma}^{+}$&2.38&2.38&2.38&2.38\\
$3_{\gamma}^{+}$&3.35&2.60&3.24&3.32\\
$4_{\gamma}^{+}$&4.42&2.87&4.28&4.44\\
$5_{\gamma}^{+}$&5.56&3.16&5.48&5.75\\
$6_{\gamma}^{+}$&7.12&3.48&6.81&7.20\\
$7_{\gamma}^{+}$&8.32&3.84&8.27&8.81\\
$8_{\gamma}^{+}$&10.37&4.22&9.86&10.54\\
$10_{\gamma}^{+}$&14.06&5.07&13.41&14.40\\
\hline
\end{tabular}
\caption{The same as in Table II but for $^{192}$Os}
\end{table}

\begin{table}
\vspace{0.5cm}
\begin{tabular}{|c|cccccc|}
\hline
Transition&Exp.&X(5)&\multicolumn{2}{c}{ISW}&\multicolumn{2}{c|}{D}\\
\hline
$2_{g}^{+}\rightarrow 0_{g}^{+}$&0.424&0.424&0.424&0.424&0.424&0.424\\
$4_{g}^{+}\rightarrow 2_{g}^{+}$&0.497&0.678&0.678&0.656&0.726&0.673\\
$6_{g}^{+}\rightarrow 4_{g}^{+}$&0.660&0.841&0.840&0.787&0.981&0.836\\
$8_{g}^{+}\rightarrow 6_{g}^{+}$&0.754&0.966&0.964&0.879&1.23&0.966\\
$10_{g}^{+}\rightarrow 8_{g}^{+}$&0.688&1.060&1.06&0.947&1.45&1.06\\
\hline
$4_{\gamma}^{+}\rightarrow 2_{\gamma}^{+}$&0.298&0.269&0.275&0.274&0.302&0.288\\
$6_{\gamma}^{+}\rightarrow 4_{\gamma}^{+}$&0.336&0.611&0.610&0.590&0.730&0.643\\
$8_{\gamma}^{+}\rightarrow 6_{\gamma}^{+}$&0.314&0.823&0.807&0.761&1.05&0.858\\
\hline
$2_{\gamma}^{+}\rightarrow 0_{g}^{+}$&0.037&0.012&0.009&0.037&0.009&0.037\\
$2_{\gamma}^{+}\rightarrow 2_{g}^{+}$&0.303&0.019&0.014&0.057&0.015&0.062\\
$4_{g}^{+}\rightarrow 2_{\gamma}^{+}$&0.014&0.001&0.006&0.026&0.007&0.029\\
$4_{\gamma}^{+}\rightarrow 2_{g}^{+}$&0.002&0.008&0.006&0.026&0.007&0.028\\
$4_{\gamma}^{+}\rightarrow 4_{g}^{+}$&0.203&0.027&0.020&0.084&0.023&0.102\\
$6_{g}^{+}\rightarrow 4_{\gamma}^{+}$&0.012&0.006&0.006&0.025&0.007&0.032\\
$6_{\gamma}^{+}\rightarrow 4_{g}^{+}$&0.0004&0.006&0.006&0.025&0.007&0.031\\
$6_{\gamma}^{+}\rightarrow 6_{g}^{+}$&0.171&0.023&0.023&0.100&0.029&0.138\\
\hline
\end{tabular}
\caption{The same as in Table III but for $^{192}$Os. The units for the B(E2) values are $e^2b^2$.}
\end{table}

The results for $^{192}$Os are presented in Tables VII and VIII. Therein, we give also the experimental data and the theoretical results yielded by the $X(5)$ formalism.
First we note two specific features for this nucleus: i) $E_{4^+}/E_{2^+}=2.82$, which recommends it as a good candidate for the $X(5)$ symmetry and ii) $E_{2^+_{\gamma}}< E_{0^+_{\beta}}$. This inequality characterizes the gamma unstable nuclei.
 From Table VII we remark the very good description of energy levels by the $D$ formalism.
Also, from there it results that the $X(5)$ formalism fails to describe the excitation energies in the gamma band.
 Concerning the E2 transitions the data from Table VIII show that the ISW formalism with a harmonic transition operator yields results very close to the ones produced with the $X(5)$ formalism. However including an anharmonic term in the expression of the transition operator one obtains a slightly better agreement with the experimental data. 

Concluding the numerical analysis of this Section  we may say that the formalism presented here  has not only the merit of removing two drawbacks of the previous X(5) model (the functions in $\gamma$ are not periodic while the Hamiltonian is not Hermitian.) but also provides  a better quantitative description.

\section{Conclusions}
Here we shall summarize the main results presented in the present paper.
Starting with the differential equation for $\beta$ and $\gamma$ variables, involving also the Euler angles, provided by the liquid drop model, one may define analytically solvable equations for both deformation variables.  This is possible under certain circumstances which are described in details in Sections II and III. The first model which achieved this situation corresponds to the so called $X(5)$ symmetry and has the merit of describing in a very simple fashion the properties of the critical point in shape phase transition. Two specific features are considered as drawbacks of the model: 1)the function describing the variable $\gamma$ are non-periodic functions and moreover are normalized to unity on a non-bounded interval. These two properties conflict the symmetry properties required by the starting liquid drop Hamiltonian. 2)In particular if we preserve the integration measure for $\gamma$ to be $|\sin3\gamma|d\gamma$ then the Hamiltonian depending on $\gamma$ is not Hermitian.

The scope of this paper was to remove these two drawbacks. The solution offered here is to use in the $\gamma$ space a Hamiltonian which has the
 spheroidal functions as eigenstates. These functions are, indeed, periodic in the interval $[0,2\pi]$. Moreover, the model Hamiltonian is Hermitian  with respect to the integration measure
$|\sin3\gamma|d\gamma$. Another solution  would be the use of a $\gamma$- Hamiltonian which admits the Mathiew functions as eigenfunctions. However, applications based on this solution are postponed for another publication. We have proved that both solutions lead  the X(5) model in the limit of $|\gamma|$ small.
As regards the $\beta$ variable, the description is performed with  the generalized Laguerre functions
which are solutions of a Schr\"{o}dinger equation involving the Davidson's potential. A particular case of this description 
is the situation when the centrifugal terms is vanishing, i.e. $\beta_0=0$. Of course, that is the oscillator potential in $\beta$.

It is inferred that each of the solvable models in the $\beta$ variable, which have been previously used as $E(5)$ models, may be associated to the spheroidal function description by separating the term proportional in $1/\beta^2$ and not depending on $\gamma$ to renormalize the equation in $\beta$. Two examples are given  in this paper where  the Davidson potential and an infinite square well potential are used. Associations of other $\beta$ potentials and the spheroidal function approach are presently under our consideration.

The numerical applications to $^{150}$Nd, $^{154}$Gd and $^{192}$Os reveal a good agreement with experimental data.
Moreover, the comparison with the results yielded by X(5) calculations suggest that the present approach provides a slightly better quantitative description of the data. For most of the data presented here the results of the present approach are close to those obtained with the X(5) formalism. However, this is not a surprising feature if we recall that the $\gamma$-Hamiltonian
has been derived through a $sin3\gamma$ expansion approach. If we consider higher angular momentum states in the three major bands, the deviations of  energies predicted by the X(5) calculations from the corresponding experimental data are larger than those obtained with the present formalism. One notes that the gamma band energies are better described by the present approach. As a matter of fact the X(5) calculations fail to reproduce the excitation energies in the gamma band of $^{192}$Os.  In the case of $^{154}$Gd we compared the theoretical results of this paper with those obtained with two versions of the Coherent State Model. The qualities of the agreement with the experimental data obtained with the three sets of calculations are comparable with each other. The essential difference is that CSM works quite well
for all even $Gd$ isotopes while the $X(5)$-type models work only for the critical point of the shape phase transition.

Perhaps this is the prize we have to pay by endorsing the variable separability. The model beauty cannot substitute 
the virtue of the variable mixing to account for details specific for one phase or another.

\clearpage
\section{Appendix A}
\label{sec:levelA}
In order to help the reader to check the expressions given in the text, here we give some intermediate results concerning
the partial expansions in $\gamma$, used in deriving the fourth order expansion for the Hamiltonian considered in Section II. Thus, the useful expansions are listed below:
\begin{eqnarray}
\frac{1}{\sin^2(\gamma-\frac{2\pi}{3})}&=&\frac{4}{3}-\frac{8\gamma}{3\sqrt{3}}+\frac{8\gamma^2}{3}-\frac{16\gamma^3}{3\sqrt{3}}+\frac{104\gamma^4}{27}+{\cal O}[\gamma]^5,\nonumber\\
\frac{1}{\sin^2(\gamma+\frac{4\pi}{3})}&=&\frac{4}{3}+\frac{8\gamma}{3\sqrt{3}}+\frac{8\gamma^2}{3}+\frac{16\gamma^3}{3\sqrt{3}}+\frac{104\gamma^4}{27}+{\cal O}[\gamma]^5,\nonumber\\
\frac{1}{\sin^2\gamma}&=&\frac{1}{\gamma^2}+\frac{1}{3}+\frac{\gamma^2}{15}+\frac{2\gamma^4}{189}+{\cal O}[\gamma]^5,\nonumber\\
\frac{1}{\sin^2(3\gamma)}&=&\frac{1}{9\gamma^2}+\frac{1}{3}+\frac{3\gamma^2}{5}+\frac{6\gamma^4}{7}+{\cal O}[\gamma]^5.
\end{eqnarray}
In order to perform the expansion around $\pi/6$ one needs the following expansions in terms
of $y=|\gamma-\pi/6|$:
\begin{eqnarray}
\frac{1}{\sin^2(\gamma-\frac{2\pi}{3})}&=&1+y^2+\frac{2y^4}{3}+{\cal O}[y]^5,\nonumber\\
\frac{1}{\sin^2(\gamma+\frac{4\pi}{3})}&=&4+8\sqrt{3}y+40y^2+\frac{176y^3}{\sqrt{3}}+\frac{728y^4}{3}+{\cal O}[y]^5,\nonumber\\
\frac{1}{\sin^2\gamma}&=&4-8\sqrt{3}y+40y^2-\frac{176y^3}{\sqrt{3}}+\frac{728y^4}{3}+{\cal O}[y]^5,\nonumber\\
\frac{1}{\sin^2(3\gamma)}&=&1+9y^2+54y^4+{\cal O}[y]^5,\nonumber\\
\end{eqnarray}

\renewcommand{\theequation}{B.\arabic{equation}}
\setcounter{equation}{0}

\section{Appendix B}
\label{sec:levelB}
In the rotational term:
\begin{equation}
W(\gamma, Q)={\frac{1}{4}}\sum_{k=1}^{3}{\frac{1}{\sin ^{2}{(\gamma -{\frac{2\pi }{3}}k)
}}}Q_{k}^{2},  \label{22}
\end{equation}
we shall consider the identity:
\begin{equation}
{\frac{1}{4}}\sum_{k=1}^{3}{\frac{1}{\sin ^{2}{(\gamma -{\frac{2\pi }{3}}k)}}}=\frac{9}{\sin^2(3\gamma)}.  
\end{equation}
The term $W(\gamma,Q)$ acquires a more convenient form:
\begin{eqnarray}
W(\gamma,Q)&=&\frac{1}{8}\left[\frac{9}{\sin^2(3\gamma)}-\frac{1}{\sin^2\gamma}\right](Q_1^2+Q_2^2+Q_3^2)+
\frac{3}{8}\left[-\frac{3}{\sin^2(3\gamma)}+\frac{1}{\sin^2\gamma}\right]Q_3^2\nonumber\\
&+&\frac{1}{8}\left[\frac{1}{\sin^2(\gamma-\frac{1\pi}{3})}-\frac{1}{\sin^2(\gamma-\frac{4\pi}{3})}\right](Q_1^2-Q_2^2)~.
\label{B3}
\end{eqnarray}
In the regime of $|\gamma|\ll 1$ one uses the expansions
\begin{eqnarray}
\frac{9}{4\sin^23\gamma}&=&\frac{1}{4\gamma^2}+\frac{3}{4}+\frac{27\gamma^2}{20}+{\cal O}(\gamma^3),\nonumber\\
\frac{1}{\sin^2\gamma} &=&\frac{1}{\gamma^2}+\frac{1}{3}+\frac{\gamma^2}{15}+{\cal O}(\gamma^3),\nonumber\\
\cos3\gamma&=&1-\frac{9\gamma^2}{2}+{\cal O}(\gamma^3),\nonumber\\
\cos^23\gamma &=&1-9\gamma^2+{\cal O}(\gamma^3),
\end{eqnarray}
in connection with the expression \ref{B3} of $W(\gamma,Q)$. The result is:
\begin{eqnarray}
W(\gamma,Q)&=&\left(\frac{1}{3}+\frac{2}{3}\gamma^2\right)(Q_1^2+Q_2^2+Q_3^2)+\left(\frac{1}{4\sin^2\gamma}-\frac{1}{3}-\frac{2}{3}\gamma^2\right)Q_3^2
\nonumber\\
&+&\frac{2}{3\sqrt{3}}(Q_2^2-Q_1^2)+{\cal O}(\gamma^3).
\end{eqnarray}
Inserting this expression into the Hamiltonian 
\begin{equation}
\tilde{H}-E=\frac{\partial^2}{\partial \gamma^2}+\frac{9}{4}\left[1+\frac{1}{\sin^23\gamma}\right]-U(\gamma)-W(\gamma,Q),
\end{equation}
and averaging the result with an axially symmetric rotational state, for which  $\langle Q_1\rangle =\langle Q_2\rangle$, and denoting the result by $H_0$, one obtains:
\begin{equation}
H_0=\frac{\partial^2}{\partial \gamma^2}+\frac{9}{4}-U(\gamma)-V(\gamma),
\end{equation}
\begin{equation}
V(\gamma)=\frac{9}{8\sin^2(3\gamma)}\left[L(L+1)-q_3^2-2\right]-\frac{1}{8\sin^2\gamma}\left[L(L+1)-3q_3^2\right].
\end{equation}
Here we used the notation $q_k=\langle Q_k\rangle$.
If the averaging is performed with a Wigner function $D^L_{MK}$, the result for $V(\gamma)$ would be:
\begin{equation}
V(\gamma)=\frac{1}{4\sin^2\gamma}(K^2-1)+\frac{1}{8}\left[\frac{9}{\sin^2(3\gamma)}-\frac{1}{\sin^2\gamma}\right]\left[L(L+1)-2-K^2\right].
\label{Vdega}
\end{equation}
In the limit of $|\gamma|\ll 1$ the above expression of $V(\gamma)$ can be expanded in powers of $\gamma$. The second order expansion is:
\begin{equation}
V(\gamma)=\frac{1}{12}\left(\frac{1}{3\gamma^2}+1+\frac{1}{5}\gamma^2\right)(K^2-1)+\frac{1}{3}\left[L(L+1)-K^2-2\right](1+2\gamma^2)+\mathcal{O}(\gamma^3).
\label{ga2}
\end{equation}
On the other hand making use of the expansion
\begin{equation}
\frac{9}{\sin^2 3\gamma}=\frac{1}{\sin^2\gamma}+\frac{8}{3}(1+2\sin^2\gamma)+\mathcal{O}(\gamma^3),
\end{equation}
one obtains the following second order expansion in $\sin\gamma$:
\begin{equation}
V(\gamma)=\frac{K^2-1}{4\sin^2\gamma}+\frac{1}{3}\left[L(L+1)-K^2-2\right](1+2\sin^2\gamma)+\mathcal{O}(\gamma^3).
\label{vsin}
\end{equation}
From this expression one obtains immediately the expansion in $\sin(3\gamma)$
\begin{eqnarray}
V(\gamma)&=&\frac{9(K^2-1)}{4\sin^23\gamma}
+\frac{1}{3}\left[L(L+1)-K^2-2\right]\left(1+\frac{2}{9}\sin^23\gamma\right)+\mathcal{O}(\gamma^3).
\label{sin3ga}
\end{eqnarray}
The three expansions for $V(\gamma)$ are useful to study different representations for the wave function in $\gamma$. Thus, inserting (\ref{ga2}) in
Eq.(B.7) one obtains a differential equation for the Laguerre functions. Here, all corrections in $\gamma^2$ were included.
Using the expansion in $\sin\gamma$ the differential equation for the variable $\gamma$ becomes an equation for the spheroidal function after the change of variable $x=cos\gamma$ is performed. 
Finally, we mention that the use of Eq.(\ref{sin3ga}) leads to a differential equation for the spheroidal functions in the variable $x=\cos (3\gamma)$.

\renewcommand{\theequation}{C.\arabic{equation}}
\setcounter{equation}{0}
\section{Appendix C}
\label{sec:levelC}
The renormalization of the equation in $\beta$ due to the terms coming form the rotational term, is based on the expansion
in $\sin3\gamma$ given by Eq. (B.13). Multiplying $V(\gamma)$ given by the quoted equation with $\frac{1}{\beta^2}$ one obtains terms which depend on $\gamma$ and terms which do not depend on this variable. The latter terms are:
\begin{equation}
R= \frac{1}{3\beta^2}\left[L(L+1)-K^2-2\right]. 
\label{Rterm}
\end{equation}
Our option for renormalization is based on the approximation:
\begin{equation}
R=\frac{1}{3\beta^2}L(L+1)-\frac{1}{3\langle\beta^2\rangle}\left[K^2+2\right]
\end{equation}
In this way the indices for the generalized Laguerre functions, if we use the Davidson potential, or of Bessel function if we use an infinite square well $\gamma$ potential are those used in the present paper and Ref.\cite{Bona3}.
 If the separation of the two types of terms is achieved differently:
 \begin{equation}
R=\frac{1}{3\beta^2}\left[L(L+1)-2\right]-\frac{1}{3\langle\beta^2\rangle} K^2
\end{equation}
then the irrational indices for Laguerre and Bessel functions are:
\begin{eqnarray}
p&=&-\frac{3}{2}+\sqrt{\frac{1}{3}\left(L+\frac{1}{2}\right)^2+\frac{3}{2}+\beta_0^4},\nonumber\\
s&=&-\frac{3}{2}+\sqrt{\frac{1}{3}\left(L+\frac{1}{2}\right)^2+\frac{3}{2}}.
\end{eqnarray}
Note that in the case the $\beta$ potential is a harmonic oscillator and the $\gamma$ potential is ignored, the separation
of the $\beta$ variable takes place in a natural manner, the $\gamma$ and Euler angles depending terms being just the Casimir operator of the group $R(5)$. Due to this separation the $\beta$ wave function is not depending on the quantum number $K$.
Here such a simple picture does not hold any longer and a $K$ depending term may renormalize the equation for $\beta$. Therefore, it becomes meaningful to consider the full term R (\ref{Rterm}) for the renormalization purpose.
In this case the two irrational indices for the generalized Laguerre function and the Bessel function are:  
\begin{eqnarray}
p&=&-\frac{3}{2}+\sqrt{\frac{1}{3}\left(L+\frac{1}{2}\right)^2 -\frac{K^2}{3}+\frac{3}{2}+\beta_0^4},\nonumber\\
s&=&-\frac{3}{2}+\sqrt{\frac{1}{3}\left(L+\frac{1}{2}\right)^2-\frac{K^2}{3}+\frac{3}{2}}.
\label{psis}
\end{eqnarray}
The difference between the index s given by Eq.(\ref{psis}) and that used in Ref.\cite{Bij} is caused by the fact the therein
the term $\frac{2}{3\beta^2}$ from Eq.(\ref{Rterm}) is not used for renormalizing the equation for $\beta$.


\begin{thebibliography}{99}

\bibitem{Bohr}
A. Bohr, Mat. Fys. Medd. Dan. Vid. Selsk. {\bf 26} (1952) no.14;
A.Bohr and B.Mottelson, Mat. Fys. Medd. Dan. Vid. Selsk. {\bf 27} (1953) no. 16

\bibitem{GrFa}A. Faessler and W. Greiner, Z. Phys. {\bf 168} (1962) 425; {\bf 170} (1962) 105; {\bf 177} (1964) 190; A. Faessler, W. Greiner and R. Sheline, Nucl. Phys. {\bf 70} (1965) 33.

\bibitem{Gneus} G. Gneus, U. Mosel and W. Greiner, Phys. Lett. {\bf 30 B} (1969) 397.

\bibitem{Hess}P. Hess, J. Maruhn and W. Greiner, Phys. Rev. {\bf C23} (1981) 2335; J. Phys. G, {\bf 7} (1981) 737.

\bibitem{Rad1}A. A. Raduta, V. Ceausescu, A. Gheorghe and R. M. Dreizler, Phys. Lett. {\bf 99B} (1981) 444; Nucll. Phys. {\bf A381} (1982) 253.

\bibitem{Rad2}A. A. Raduta, V. Ceausescu and A. Faessler, Phys. Rev. {\bf C36} (1987) 2111.

\bibitem{Rad3}A. A. Raduta, C. Lima and A. Faessler, Z. Phys. {\bf A 313} (1983) 69.

\bibitem{Rad4}A. A. Raduta, Al. H. Raduta and A. Faessler, Phys. Rev. {\bf C55} (1997) 1747; 
A. A. Raduta, D. Ionescu and A. Faessler, Phys. Rev. {\bf C65} (2002) 064322.

\bibitem{Rad5}A. A. Raduta, in Recent Res. Devel. Nuclear Phys.,1 (2004):1-70, ISBN:81-7895-124-X.

\bibitem{Jean}L. Wilets and M. Jean, Phys. Rev. {\bf 102} (1956) 788.

\bibitem{Filip} A. S. Davydov and G. F. Filippov, Nucl. Phys. {\bf 8} (1958) 788.

\bibitem{Iache}A. Arima and F. Iachello, Ann. Phys.(N.Y.) {\bf 99} (1976) 253;
{\bf 123} (1979) 468.

\bibitem{Iache1}F. Iachello and A. Arima, The Interacting Boson Model (Cambridge University Press, Cambridge, England, 1987).

\bibitem{Cast}R. F. Casten, in Interacting Bose-Fermi Systems in Nuclei, edited by F. Iachello (Plenum, New York, 1981), p. 1.

\bibitem{Row1} D. J. Rowe, Nucl. Phys. {\bf A 745} (2004) 47.

\bibitem{Row2} P. S. Turner, D. J. Rowe, Nucl. Phys. {\bf A 756} (2005) 333.

\bibitem{Row1} G. Rosensteel, D. J. Rowe, Nucl. Phys. {\bf A 759} (2005) 92.

\bibitem{Gino}J. H. Ginocchio and M. W. Kirson, Phys. Rev. Lett.{\bf 44}(1980) 1744.

\bibitem{Diep} A. E. L. Dieperink, O. Scholten and F. Iachello, Phys. Rev. Lett.{\bf 44} (1980) 1747.

\bibitem{Iache2} F. Iachello, Phys. Rev. Lett. {\bf 85} (2000) 3580.

\bibitem{Iache9} F. Iachello, Phys. Rev. Lett. {\bf 87} (2001) 052502.

\bibitem{Zam} R. F. Casten and N. V. Zamfir, Phys. Rev. Lett. {\bf 85} (2000) 3584.

\bibitem{Zam1} R. F. Casten and N. V. Zamfir, Phys. Rev. Lett. {\bf 87} (2001) 052503.

\bibitem{Clar}R. M. Clark, M. Cromaz, M. A. Deleplanque, M. Descovich, R. M. Diamond, P. Fallon,
I. Y. Lee, A. O. Macchiavelli, H. Mahmud, E. Rodriguez-Vieitez, F. S. Stephens and D. Ward,
Phys. Rev. {\bf C 69} (2004) 064322.

\bibitem{Zam2}N. V. Zamfir et al., Phys. Rev. {\bf C65} (2002) 044325.

\bibitem{Da-li}Da-li Zhang and Yu-xin Liu, Phys. Rev. {\bf C65} (2002) 057301.

\bibitem{Fort} L. Fortunato and A. Vitturi, Jour. Phys. G:Nucl. Part. Phys.
{\bf 29} (2003) 1341.

\bibitem{Bona} Dennis Bonatsos, D. Lenis, N. Minkov, D. Petrellis, P. P. Raychev, P. A. Terziev,  Phys. Lett. {\bf B 584} (2004) 40.

\bibitem{Davi} P. M. Davidson, Proc. R. Soc. {\bf 135} (1932) 459.

\bibitem{Rad05} A. A. Raduta, A. Gheorghe and A. Faessler, J. Phys. G:Nucl. Part. Phys.,{\bf 31} (2005) 337.

\bibitem{Rad06} A. A. Raduta, F. D. Aaron and I. I. Ursu, Nucl. Phys. {\bf A 772} (2006) 20.

\bibitem{Caprio07} M. A. Caprio, P. Cejnar and F. Iachello, arXiv:0707.0325v1[quant-ph], Annals of Physics NY, 323 (2008) 1106.

\bibitem{Fortu2}L. Fortunato, arXiv:Nucl-Th/0411087v1; Eur. Phys. J. {\bf A 26} (2005) s01.

\bibitem{Ghe78}A. Gheorghe, A. A. Raduta and V. Ceausescu, Nucl. Phys. {\bf A 296} (1978) 228.

\bibitem{Rad78} A. A. Raduta, A. Gheorghe and V. Ceausescu, Nucl. Phys. {\bf A311} (1978) 118.


\bibitem{Rad98} A. Gheorghe, A. A. Raduta and V. Ceausescu, Nucl. Phys. {\bf A637}
(1998) 201.

\bibitem{Moshy} E. Chac\'{o}n, M. Moshynsky and R. T. Sharp, J. Math. Phys. {\bf 17} (1976) 668.

\bibitem{Willi} T. M. Corrigan, F. J. Margetan and S. A. Williams, Phys. Rev. {\bf C 4} (1976) 2279.

\bibitem{Row} D. J. Rowe, Nucl. Phys. {\bf A 735} (2004) 372.

\bibitem{Bes} D. B\'{e}s, Nucl. Phys. {\bf 210} (1959) 373.

\bibitem{Rad07} A. Gheorghe, A. A. Raduta and A. Faessler, Phys. Lett. {\bf B 648}, 171, (2007).

\bibitem{Dav} P. M. Davidson, Proc. R. Soc. {\bf 135} (1932) 459.

\bibitem{Elli}J. P. Elliot, J. A. Evans, P. Park, Phys. Lett. {\bf B 169} (1986) 309.

\bibitem{Row} D. J. Rowe, C. Bahri, J. Phys. {\bf A 31} (1998) 4947

\bibitem{Abra}M. Abramowitz,  and I. A. Stegun, (Eds.), Handbook of Mathematical Functions with Formulas,Graphs, and Mathematical Tables, 9th printing. New York: Dover,1972, pp. 751-759.

\bibitem{Baerd} S. De Baerdenmacker, L. Fortunato, V. Hellemans, K. Heyde, Nucl. Phys. {\bf A 769}
(2006) 16.

\bibitem{Kruc} R. Kr\"{u}cken {\it et al.} Phys. Rev. Lett. {\bf 88} (2002) 232501.

\bibitem{Mateo} E. der Mateosian, J. K. Tuli, NDS {\bf 75} (1995) 827.

\bibitem{Clar1} R. M. Clark {\it et al.}, Phys. Rev. {\bf C68} (2003) 037301.

\bibitem{Lipa} P. O. Lipas {\it et al} Phys. Scr. 7 (1983) 8.

\bibitem{Giri} C. Girit, W. D. Hamilton and C. A. Katfas, J. Phys. G: Nucl. Part. Phys.{\bf 9} (1983) 797. 

\bibitem{Rei} C. W. Reich, R. G. Helmer, NDS {\bf 85} (1998) 171.

\bibitem{Rad055} A. A. Raduta and Amand Faessler, J. Phys. G: Nucl. Part. Phys. {\bf 31} (2005) 873.

\bibitem{Allm} J. M. Allmond {\it et al.} Phys. Rev. {\bf C 78} (2008) 014302.

\bibitem{Cora} Coral M. Baglin, NDS {\bf 84} (1998) 717.

\bibitem{Lala}G. A. Lalazissis, S. Raman and P. Ring, At. Data Nucl. Data Tables, {\bf 71} (1999) 1.

\bibitem{Bij}R. Bijker, R. F. Casten, N. V. Zamfir, E. A. McCutchan, Phys. Rev. {\bf C 68} (2003) 064304.

\bibitem{Caprio} M. A. Caprio, Phys. Rev. {\bf C69} (2004) 044307.

\bibitem{Bona3} D. Bonatsos, D. Lenis, E. A. McCutchan, D. Petrellis, I. Yigitoglu,
Phys. Lett. {\bf B649} (2007) 394.

\bibitem{Nils} S. G. Nilsson, Mat. Fys. Medd. K. Dan. Vid. Selsk.{\bf 29}, no. 16, (1955) 1.

\bibitem{RadIud} A. A. Raduta, D. S. Delion and N. Lo Iudice, Nucl. Phys.
{\bf A 564} (1993) 185.


\bibitem{Palma} G. Palma and U. Raff Am. J. Phys. {\bf 71} (2003) 956.

\end{thebibliography}
\end{document}